\documentclass[letterpaper,authoryear]{elsarticle}
\usepackage{times}
\usepackage{float}
\usepackage{amsmath}
\usepackage{epsfig}
\usepackage{lineno}
\usepackage{subcaption}
\usepackage[breaklinks,colorlinks,citecolor=blue]{hyperref} 
\usepackage{cleveref}       
\usepackage[all]{hypcap}
\usepackage{outlines}
\usepackage{setspace}
\begin{document}

\begin{frontmatter}

\title{Climate Diversity in the Solar-Like Habitable Zone due to Varying Background Gas Pressure}
\author[daa,cps]{Adiv Paradise\corref{cor1}}
\ead{paradise@astro.utoronto.ca}
\author[phys,eco]{Bo Lin Fan}
\author[daa,pag]{Kristen Menou}
\author[phys]{Christopher Lee}

\cortext[cor1]{Corresponding author}
\address[daa]{David A. Dunlap Department of Astronomy \& Astrophysics, University of Toronto, St. George, Toronto, ON M5S 3H4, Canada}
\address[cps]{Centre for Planetary Sciences, Department of Physical and Environmental Sciences, University of Toronto, Scarborough, ON M1C 1A4, Canada}
\address[pag]{Physics \& Astrophysics Group, Department of Physical and Environmental Sciences, University of Toronto, Scarborough, ON M1C 1A4, Canada} 
\address[phys]{Department of Physics, University of Toronto, St. George, Toronto, Ontario, M5S 1A7, Canada}
\address[eco]{Department of Ecology \& Evolutionary Biology, University of Toronto, St. George, Toronto, Ontario, M5S 3B2}

% \affil[*]{Corresponding author email: paradise@astro.utoronto.ca}


\journal{Icarus}

% \maketitle

\begin{abstract}
% 
% With the discovery of thousands of exoplanets, including a growing number of Earth-sized exoplanets at Earth-like instellations, there has been a growing emphasis on understanding their atmospheres and climates. Theoretical efforts usually assume either a CO$_2$-dominated atmosphere, or an N$_2$-dominated atmosphere, like Earth's, and in the latter case, the usual assumption is a surface pressure similar to Earth's. There have to date been only limited efforts to understand how varying the amount of N$_2$ would affect the climate, as large, 3D climate models are resource-intensive, N$_2$ is difficult to observe with transit spectroscopy due to a lack of absorption lines in the visible or infrared, and it is often assumed that N$_2$ does not directly affect the planet's energy budget. 

A large number of studies have responded to the growing body of confirmed terrestrial habitable zone exoplanets by presenting models of various possible climates. However, the impact of the partial pressure of background gases such as N$_2$ has not yet been well-explored, despite the abundance of N$_2$ in Earth's atmosphere and the lack of constraints on its typical abundance in terrestrial planet atmospheres. We use PlaSim, a fast 3D climate model, to simulate many hundreds of climates around Sun-like stars with varying N$_2$ partial pressures, instellations, and surface characteristics to identify the impact of the background gas partial pressure on the climate. We find that the climate's response is nonlinear and highly sensitive to the background gas partial pressure. We identify pressure broadening of greenhouse gas (such as CO$_2$ and H$_2$O) absorption lines, amplification of warming or cooling by the water vapor greenhouse positive feedback, heat transport efficiency, and cooling through Rayleigh scattering as the dominant competing mechanisms that determine the equilibrium climate for a given N$_2$ partial pressure. Finally, we show that different amounts of N$_2$ should have a significant effect on broadband reflected light observations of terrestrial exoplanets around Sun-like stars.
% \begin{outline}
% \1 We've discovered a lot of planets, and transmission spectroscopy is one of the dominant ways of characterizing their atmospheres.
% \1 Transmission spectroscopy has a hard time measuring the mass of background gases like N$_2$ that don't absorb in the visible or infrared.
% \1 Earth's atmosphere is mostly background gases.
% \1 We don't actually know what the range of likely background gas masses on terrestrial planets is.
% \1 We used PlaSim to see what impact N$_2$ has on the climate by varying pN$_2$ across instellation.
% \1 We find pN$_2$ has a large nonlinear effect on the climate, and constraining background gas contributions is essential to characterizing the climates of terrestrial planets.
% \end{outline}
\end{abstract}
\end{frontmatter}

% \linenumbers
\section{Introduction}

The typical starting assumption for climate models of Earth-like planets is either a CO$_2$-dominated atmosphere like that on Mars and Venus, or an Earth-like atmosphere, with approximately 1 bar of N$_2$, trace CO$_2$, and a small amount of water vapor determined by evaporation and precipitation \citep[e.g][]{Shields2015,Turbet2016,Boutle2017,Wolf2017}. These are not unreasonable assumptions; our Solar System presents limited opportunities for in-depth studies of habitable zone terrestrial planets with atmospheres. Additionally, sophisticated climate models are computationally resource-intensive, limiting the number of models that can be included in sensitivity studies \citep[e.g.][]{Way2017}, and other parameters such as pCO$_2$, rotation rate, eccentricity, and obliquity all have strong effects and known dynamical and geophysical constraints \citep[e.g.][]{Manabe1975,Williams1997,Shields2015,Haqq-Misra2018}. pCO$_2$ in particular has well-studied geochemical constraints\citep{walker81,Berner2004,pierrehumbertbook,Valencia2018,Nakayama2019}, which makes it an appealing parameter to vary \citep[e.g.][]{Kopparapu2013,Turbet2016,Paradise2017}. While there has been considerable recent progress in understanding N$_2$ geochemistry \citep{Wordsworth2015}, N$_2$ is nonetheless less well-understood, and is very difficult to observe with transit spectroscopy due to a lack of absorption lines at visible or infrared wavelengths \citep{Lofthus1977,Benneke2012}. We focus on N$_2$ in this study, but this observational problem extends to all spectrally-inactive background gases, and our results are therefore likely to apply to them as well. Because N$_2$ is chemically mostly-inert and lacks absorption lines and therefore cannot directly act as a greenhouse gas, its dominant role in the climate is often assumed to be as a simple background gas pressure that keeps water liquid at Earth-like temperatures, enables heat transport, and causes pressure broadening of the CO$_2$ and H$_2$O absorption lines \citep{Kopparapu2014,Wordsworth2015,Olson2018,Ramirez2018a}. 

While a number of authors have probed the role of background gas pressure, existing work has been limited by simplifying assumptions and computational expense, such that while many pieces of the puzzle have been described, the overall role of background gases such as N$_2$ throughout the habitable zone of Sun-like stars has yet to be fully-explored in a comprehensive, systematic manner. \citet{Nakajima1992} and \citet{Koll2019} argued that increasing N$_2$ partial pressure (pN$_2$) would likely lead to cooler climates, as the atmospheric lapse rate would trend towards the dry adiabat, resulting in more efficient cooling. \citet{Kasting1993} considered how 10 bars of N$_2$ might change the limits of the habitable zone relative to 1 bar, finding little change to the inner and outer edges, but noting stabilization against water loss, which was confirmed by \citet{Koll2019}. Several studies \citep[e.g.][]{Goldblatt2009,Li2009,Charnay2013,Goldblatt2013,Kopparapu2014,Wolf2014,Chemke2016} have shown that pressure-broadening of greenhouse gases by N$_2$ should lead to a stronger greenhouse effect, which in the absence of a strengthened cooling mechanism such as increased cloud reflectivity or Rayleigh scattering should lead to net warming with increased pN$_2$. \citet{Goldblatt2009}, \citet{Goldblatt2013}, \citet{Kopparapu2014}, and \citet{Keles2018} included Rayleigh scattering, and found that at high pressures, increased scattering and reflection of shortwave radiation could overcome pressure broadening and lead to net cooling. The importance of pressure broadening at moderate surface pressures might suggest cooler climates at low pN$_2$, but \citet{Zahnle2016} noted that low surface pressures would result in less effective trapping of water vapor in the the lower atmosphere, as the tropopause is warmer with low surface pressures \citep{Wordsworth2014}. Increased stratospheric water vapor would then lead to increased total water vapor and a stronger greenhouse effect, suggesting that climates are warmer at low pN$_2$ than at higher pressures. Conversely, \citet{Kaspi2015} showed in a study of atmospheric dynamics at different pressures that weaker vertical heat transport could lead to warming at higher pressures. \citet{Vladilo2013} included the effects of heat transport, increased heat capacity, and pressure broadening in an ensemble of one-dimensional energy-balance models, and found that the habitable zone appeared broader at high pressures, due primarily to warming at low instellations and reduced water loss at high instellations. However, they increased pCO$_2$ proportionally with pN$_2$, making it difficult to isolate warming from N$_2$. \citet{Kaspi2015} and \citet{Chemke2017} modeled a range of N$_2$ partial pressures with a 3D climate model, but omitted pressure broadening and Rayleigh scattering, examining only the dynamical effects. 

In recent years, fully-coupled 3D models have been used to vary pN$_2$. \citet{Turbet2018} considered models of TRAPPIST-1e, TRAPPIST-1f, TRAPPIST-1g, and TRAPPIST-1h at various N$_2$ and CO$_2$ partial pressures, but did not vary instellation for each planet, and limited their investigation of the parameter space to the question of atmospheric collapse on the night-side of tidally-locked planets. Furthermore, because their investigation was limited to models of specific tidally-locked exoplanets around a 2500 K star, it may be difficult to extrapolate their results to fast-rotating planets around Sun-like stars. Most recently, \citet{Komacek2019} presented the results of five models computed using ExoCAM, a fully-coupled three-dimensional climate model, with pN$_2$ ranging from 0.25 to 4 bars. They found a nonlinear response, with warming up to 1 bar, and dramatic cooling at 2 and 4 bars, which they attributed to increased Rayleigh scattering. However, their models did not include CO$_2$, and they note significant disagreement with previous studies \citep{Goldblatt2009,Li2009,Charnay2013,Wolf2014,Chemke2016} that demonstrated warming through pressure broadening. The nonlinear behavior they observed was nonetheless consistent with that observed in \citet{Goldblatt2009} and \citet{Keles2018}, and consistent with studies of the runaway greenhouse threshold finding that elevated pN$_2$ could both reduce absorbed shortwave radiation as well as reduce outgoing longwave radiation \citep{Goldblatt2013,Kopparapu2014}.

Studying the climatic effects of pN$_2$ is of particular importance for understanding the climate evolution of Earth itself, particularly in the context of the ``Faint Young Sun paradox" \citep{Sagan1972}, as \citet{Goldblatt2009} proposed a higher N$_2$ partial pressure as a partial solution. However, the actual pN$_2$ during Earth's early history is a matter of debate. Fossilized raindrops suggest that as late as 2.7 Gya, Earth's pN$_2$ was less than half its current levels, and likely was not that different 3--3.5 Gya \citep{Som2016}. However, \citet{Johnson2017} showed that there was a secular increase in crustal nitrogen during the Precambrian, suggesting that pN$_2$ has been decreasing over time, supporting their finding in \citet{Johnson2015} that the bulk silicate Earth contains several bars of nitrogen. Emplacing that much nitrogen in the mantle after the hot conditions of the Hadean is difficult \citep{Wordsworth2015}, though \citet{Stueken2016} argued that biological burial and release could lead to sharper swings in pN$_2$ than would be permitted by geochemistry alone. A deeper understanding of pN$_2$'s effect on the climate would make it easier to resolve the Faint Young Sun paradox by using paleoclimate and geological pN$_2$ indicators to constrain the evolution of Earth's atmosphere. 

In this study, we use PlaSim, a fast 3D climate model, to explore the climates of Earth-sized planets in the habitable zone with pN$_2$ levels ranging from 0.1 to 10 bars. We vary instellation, synchronicity, presence of land, and initial conditions to demonstrate the range of climates permitted by variations in pN$_2$, then perform a series of experiments to thoroughly identify the mechanisms responsible for variations in climate. Finally, we explore the impacts of pN$_2$ on observations of Earth-like exoplanets.

We caution the reader before continuing that PlaSim, as will be shown and discussed in the sections that follow, is of only intermediate complexity, and lacks some of the state-of-the-art physics parameterizations necessary to accurately model climates very different from Earth's. The simplifications in the model permit extremely fast computation and the exploration of very large parameter spaces, but may limit the quantitative accuracy of the model in parts of the parameter space far from Earth conditions. Our comparison with SBDART, as described in the next section and shown in \autoref{fig:comparison}, leads us to believe that that qualitative conclusions can be made nonetheless, and that the benefits of surveying large parameters spaces quickly outweigh the loss of quantitative accuracy at the edges of the parameter space. Readers should nonetheless take caution in drawing conclusions from our results beyond the qualitative conclusions we present in this paper.

\section{Methods}
\subsection{Climate Model}
We use PlaSim, a 3D general circulation model (GCM) designed for Earth-like climates \citep{Fraedrich2005}. PlaSim uses a spectral core to solve the primitive fluid equations for pressure, temperature, divergence, vorticity, and water vapor advection. We use the model in its T21 configuration, in which the atmosphere is discretized into 32 latitudes, 64 longitudes, and 10 vertical levels. The atmosphere is coupled to a surface model, which includes a mixed-layer slab ocean, thermodynamic sea ice, and a soil model on land that allows for variable soil moisture, advection of excess water to continental margins (the model's river system), and icy or snowy surfaces. PlaSim's hydrological cycle is relatively complete, including soil storage, runoff, surface evaporation, storage as atmospheric water vapor, cloud formation, deep and shallow convection, and precipitation as either rain or snow, including re-evaporation of falling precipitation as the water passes through warmer layers. Cumulus convection is computed using a scheme similar to \citet{Kuo1965,Kuo1974}, while stratiform and cirrus clouds are computed diagnostically from precitation and relative humidity respectively, following \citet{Slingo1991}. The cloud liquid water path is computed following \citet{Kiehl1996} from the air density, height, and vertically-integrated water vapor.

Radiation is computed using a three-band model, with two shortwave bands and one longwave band. Transmissivity is computed in each band according to absorber abundances, while clouds are responsible for gray scattering based on cloud fraction and liquid water path, following \citet{Stephens1978,Stephens1984a} and \citet{Stephens1984b}. Where there are multiple fractional cloud layers, overlap is assumed to be random. Transmissivities in the shortwave scheme are computed following \citet{Lacis1974}, and transmissivities and emissivities in the longwave regime are derived from \citet{Sasamori1968}. Rayleigh scattering is computed only in the bottom layer. Pressure broadening of each absorbing species is included, and PlaSim assumes that the pressure-broadened effective abundance of each species scales linearly with local pressure \citep{Fraedrich2005}, following \citet{Sasamori1968} and \citet{StrongPlass1950}.

\subsection{Modifications to PlaSim}
PlaSim does not natively consider varying surface pressure when computing Rayleigh scattering. The scattering transmittance in the bottom layer is computed as direct and diffuse components with prescribed scattering efficiencies \citep{Fraedrich2005}:
\begin{linenomath*}
\begin{align}
T_\text{direct} &= 1-\frac{0.219}{1+0.816\mu_0} \label{eq:plasim1}\\ \nonumber \\
T_\text{diffuse} &= 0.856 \label{eq:plasim2}
\end{align}
\end{linenomath*}
where $\mu_0$ is the cosine of the solar zenith angle. We note that $T_\text{diffuse}=1-0.144$, and that 0.144 is the actual fit parameter corresponding to PlaSim's assumed scattering optical depth for diffuse radiation. To adapt this prescription for variable surface pressure, we assume that the scattering optical depth of a column of atmosphere scales mostly linearly with the mass of that column, and therefore surface pressure. According to the Beer-Lambert Law, transmittance is exponentially related to the optical depth $\tau$ \citep{beerlambert}:
\begin{linenomath*}
\begin{equation}
T = \exp(-\tau)
\end{equation}
\end{linenomath*}
Our modified prescription therefore computes the optical depth, scales it by the column mass relative to that on Earth, and then computes the transmittance:
\begin{linenomath*}
\begin{equation}
T' = \exp\left[\left(\frac{p_s}{p_{s,\oplus}}\frac{g_\oplus}{g}\right)\ln{T_0}\right]
\end{equation}
\end{linenomath*}
where $T_0$ is the prescribed transmittance for Earth (direct or diffuse), $p_s$ is the surface pressure, and $g$ is the surface gravity.

We also modified PlaSim's vertical discretization to accommodate variable surface pressures. PlaSim uses a sigma system of vertical discretization, where the vertical coordinate is dimensionless, and its value at a given level is simply the ratio of the current pressure to the surface pressure, such that a vertical coordinate of 1 indicates the surface, and 0 indicates the top of the atmosphere \citep{Fraedrich2005}. PlaSim computes its vertical levels with a fourth-order polynomial that is mostly linear between 0 and 1:
\begin{linenomath*}
\begin{align}
\sigma_{h,k} &= \frac{3}{4}\sigma_k+\frac{7}{4}\sigma_k^3-\frac{3}{2}\sigma_k^4 \label{eq:vert}\\ \nonumber \\
\sigma_k' &= \frac{1}{2}(\sigma_{h,k}+\sigma_{h,k-1})
\end{align}
\end{linenomath*}
where $\sigma_k$ is linearly-spaced between 0 and 1, $\sigma_{h,k}$ is the coordinate of the interface between model layers $k$ and $k+1$, and $\sigma_k'$ is the coordinate of layer $k$'s mid-point. This scheme is designed to give slightly higher resolution near the surface and near the tropopause, but because it starts from coordinates evenly-spaced in pressure, increasing the surface pressure results in higher pressures for all model layers. With a surface pressure of 10 bars, this means there is almost 1 bar of atmosphere in the model's top layer. This means important radiative physics can be lost in the low resolution of the top layer. To mitigate this, we simply rescale the layer interface coordinates so that the top interface always occurs at a prescribed pressure $p_\text{top}$, or $p_\text{top}/p_s$ in sigma-coordinates. We use 50 mbar for most models, ensuring that the top model layer always spans only the top 50 mbar of the atmosphere. For models with surface pressures less than 1 bar, we linearly scaled this down to 5 mbar at $p_s=0.1$ bar. 

We found in our testing that the model was not particularly sensitive to the choice of $p_\text{top}$, though choosing a value too low for a high-pressure model could affect model stability. We also experimented with doubling the vertical resolution to 20 levels, and replacing the vertical discretization scheme entirely with one that was logarithmic in pressure (linearly-spaced in altitude). We found little difference between pinning the top layer and doubling the resolution, but using a logarithmic scheme resulted in higher peak temperatures in the warmest high-pressure models, and reduced PlaSim's stability while running and required shorter timesteps. We therefore conclude that using a prescribed $p_\text{top}$ is sufficient for this study.  

\begin{figure}
\begin{center}
\includegraphics[width=5.0in]{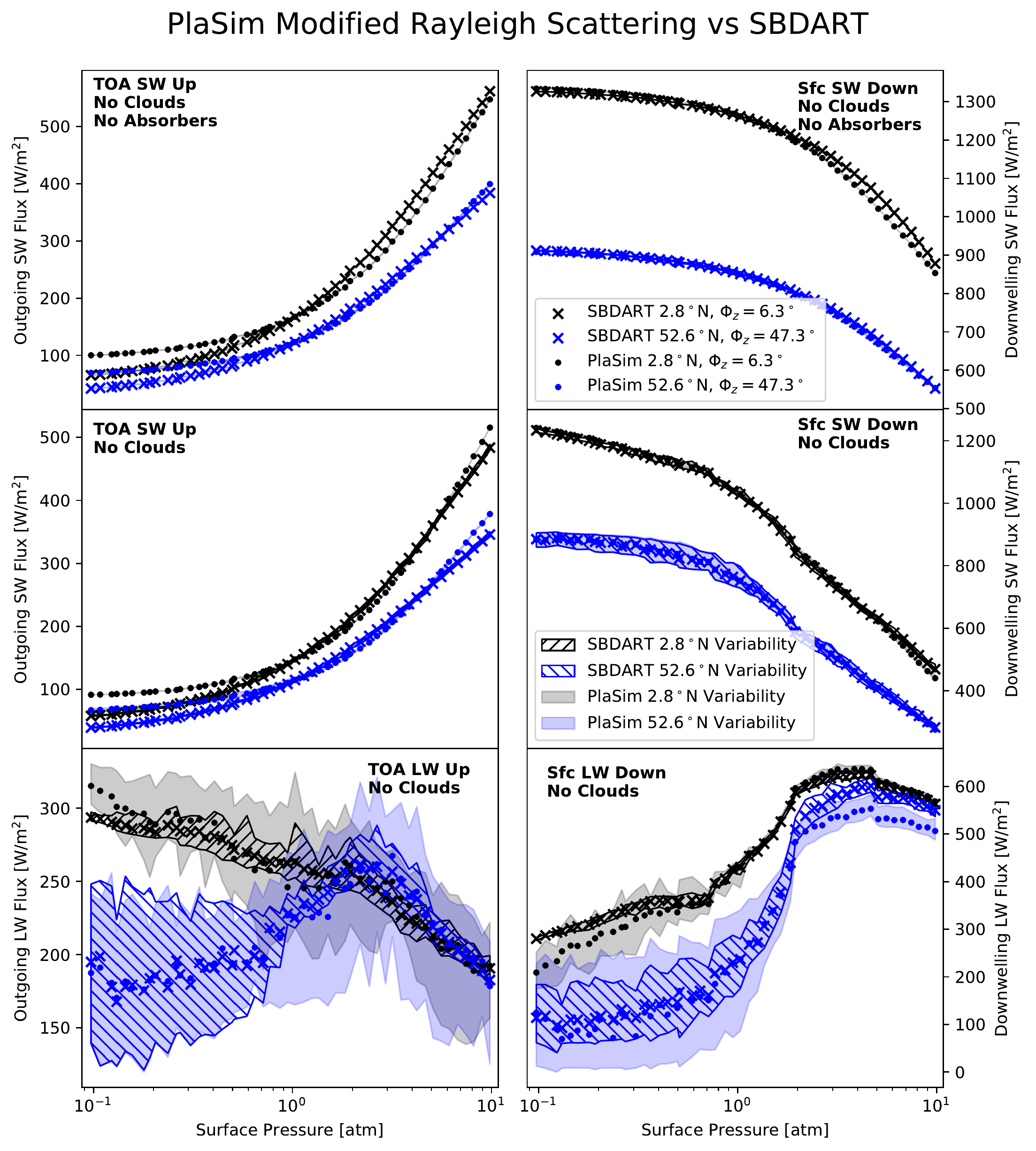}
\end{center}
\caption{Top-of-atmosphere outgoing fluxes and surface downwelling fluxes computed with PlaSim's shortwave model with our modified Rayleigh scattering and vertical discretization, compared to SBDART's fluxes, for an instellation of 1350 W/m$^2$. We ignore clouds in each test. This comparison uses as input a column of PlaSim output at an equatorial gridpoint and at a mid-latitude gridpoint from 50 equilibriated PlaSim models at 1350 W m$^{-2}$, spanning surface pressures of 0.1--10 bars. Gridpoint latitude $\theta$ and solar zenith angle $\Phi_z$ are given in the legend, and markers indicate annual average fluxes for a given column, while shaded regions represent flux variability throughout the year. We assign an ocean surface for each column to avoid the effects of sea ice or land, and we assume a flat Lambertian ocean albedo, with no dependence on the direct beam zenith angle. Our modified PlaSim radiation model shows good agreement with SBDART at both the equator and mid-latitudes for a range of surface pressures, but slightly overpredicts scattering at pressure extremes. We attribute this to PlaSim's Rayleigh scattering implementation, which only scatters in the bottom layer. PlaSim's downwelling longwave fluxes are systematically low at high pressures for mid-latitudes and at low pressures near the equator, which we attribute to the simpler broadband emissivity and absorptivity model PlaSim uses}\label{fig:comparison}
\end{figure}

To verify the validity of these modifications, we compared PlaSim's shortwave fluxes to those computed with SBDART \citep[Santa Barbara DISORT Atmospheric Radiative Transfer,][]{Ricchiazzi1998}, which computes plane-parallel radiative transfer as a function of wavelength. We used 50 PlaSim models at 1350 W m$^{-2}$ in an aquaplanet configuration, with surface pressures ranging from 0.1 to 10 bars. We used as input to SBDART a single column of output from each model, computing fluxes at 24 points throughout the year at local noon, at both an equatorial gridpoint and a mid-latitude gridpoint. We ignored clouds and sea ice, and computed broadband fluxes at each layer from 250 to 2500 nm. We then ran PlaSim's shortwave radiation model on that same column, making the same assumptions, and compared the outgoing shortwave fluxes at the top of the atmosphereand downwelling shortwave fluxes at the surface. We repeated this comparison for longwave fluxes, using 2.5--80 $\mu$m as our wavelength range in SBDART. Finally, to isolate the behavior of the different scattering schemes, we repeated the shortwave comparison for a case with no atmospheric absorbers. The results of this comparison are shown in \autoref{fig:comparison}. Our modified model agrees with SBDART in downwelling shortwave fluxes at the surface and outgoing longwave fluxes at the top of the atmosphere to within a few percent with surface pressures up to 10 bars, but we note that outgoing shortwave fluxes at the top of the atmosphere are over-predicted relative to SBDART at pressure extremes, and downwelling longwave fluxes at the surface are underpredicted for mid-latitudes at high pressures, as well as at the equator at low pressures. It is possible to reduce the disagreement in the shortwave by adjusting the constants in \autoref{eq:plasim1} and \autoref{eq:plasim2}, but this would change PlaSim's overall output when applied to the standard Earth climate. PlaSim has been used extensively for Earth-like climates \citep[e.g.][]{Fraedrich2005,Boschi2013,plasimgenie,Nowajewski2018,plasimAGU2018}, and we do not wish to make modifications that would require significant re-tuning and re-validation on the modern Earth climate. We therefore accept these disagreements as a difference in model parameterization, and conclude from the overall agreement between PlaSim and SBDART that our modified model is sufficiently accurate for this study.

Finally, we wish to note that we have only evaluted PlaSim's radiative performance with Earth-like models, and have not evaluated its performance in extremely hot and wet atmospheres, such as those explored in \citet{Goldblatt2013}. PlaSim lacks the physics necessary to model atmospheres where a volatile gas such as water is a primary constituent, instead assuming that volatiles such as water are minor components. The reader should therefore take caution before interpreting climate trends in our warmest climates as anything other than qualitative. Additional research using more sophisticated GCMs is needed to make specific climate predictions for planets with high surface pressures at high instellations.

\subsection{Modeling Strategy}\label{sec:strategy}
\subsubsection{Fast Rotators}\label{sec:strat-earth}
To assess the effect of pN$_2$ on the climate of fast-rotating planets such as Earth, we kept pCO$_2$ constant at 360 $\mu$bar, and varied pN$_2$ from 0.1 to 10 bars, and instellation from 1100 to 1550 W m$^{-2}$, computing a grid of 190 models. We assume a 24-hour rotation period in these models. PlaSim does not differentiate between background gases such as O$_2$ and N$_2$, so here we assume that the entirety of the atmosphere's background gas fraction is N$_2$, and vary pN$_2$ alone by increasing the surface pressure while reducing the CO$_2$ concentration. To limit our analysis to the role of background gas pressure, we keep the gas constant at its Earth value of 287 J/kg/K. Since here we define a `background gas' as having negligible contribution to the energy budget through visible or infrared absorption or emission lines, we do not expect the choice of a different background gas to have much impact on our results beyond changing the atmosphere's mean molecular weight, with the notable exception that H$_2$-dominated atmospheres are likely to behave very differently due to having a larger thermal scale height than a steam atmosphere \citep{Koll2019}. To isolate the dynamical effects of surface pressure and the radiative effects of N$_2$, we assume a fixed background gas mean molecular weight identical to that in Earth's atmosphere. We run each model for a minimum of 250 years until both the surface and top of the atmosphere reach a stable energy balance, which in many cases is satisfied by 250 years, but in some of the more extreme cases can take a few more centuries. We run PlaSim on 16-core nodes on our local computing cluster, which takes approximately 45 seconds per model year at the default timestep of 45 minutes. We assume Earth's modern continental distribution and topography for each model in this grid, but we also perform several experiments with aquaplanet configurations (no land; sea surface everywhere) to simplify the model and rule out the role of land surface processes. We assume that the ocean mixed layer thickness is 50 meters in all our models, and initialize each model in PlaSim's default warm-start initial conditions, in which there is no initial snow or sea ice, and a 250 K isothermal atmosphere is allowed to adjust to a fixed 288 K surface temperature before model integration begins, at which point all temperatures are allowed to evolve. In this state, the model quickly spins up to an initially-temperate climate for the instellations used, and only once sea ice has had an opportunity to begin forming is the transition into snowball possible. 

\subsubsection{Synchronous Rotators}\label{sec:strat-tl}
As the efficiency of heat transport is dependent on surface pressure, we further test its relative importance by considering the case of tidally-locked planets, which exhibit 1:1 synchronous rotation. On these planets, advection of heat from the day-side to the night-side is much more efficient than on Earth-like fast-rotating planets \citep{Checlair2017,Haqq-Misra2018}. We would therefore expect to see enhanced cooling due to heat transport at higher pN$_2$ on tidally-locked planets than on fast-rotating planets. To investigate this, we computed a grid of 483 models with PlaSim in a tidally-locked configuration, with instellations ranging from 400 to 2600 W m$^{-2}$ and surface pressures ranging from 0.1 to 12 bars. Each model had a rotation period of 12 days, and as with the fast rotators was initialized with PlaSim's default warm-start initial conditions For simplicity, each model was in an aquaplanet surface configuration. PlaSim has been used before to study tidally-locked planets \citep{Menou2013,Checlair2017,Abbot2018}, but we introduced small modifications to ensure that numerical hyper-diffusion timescales (the timescales on which PlaSim damps numerical noise introduced by its dynamical core) would use 24-hour days for unit conversions between days and seconds, rather than simply using the current length of the solar day (which is properly undefined for a tidally-locked planet). Our tests indicate that this modification does not affect the results of \citet{Menou2013,Checlair2017}, or \citet{Abbot2018}, as those models were not in a regime strongly-limited by numerical hyper-diffusion. While most tidally-locked habitable zone planets orbit low-mass stars with much redder spectra than the Sun, here we assume a solar-like input spectrum. We do not include the effects of a different input spectrum because we wish to isolate the ways in which the geometry and dynamics of tidally-locked planets affect the relative strengths of the mechanisms of action we identified in \autoref{sec:mechanisms}. This also allows us to avoid having to consider the increased role that near-infrared water absorption may play in the climates of planets around redder stars. The results of this experiment should therefore not be taken as quantitative predictions of the climate on real tidally-locked planets, but rather as a dynamical experiment. Further work will be required to investigate how a redder spectrum affects the climate's sensitivity to pN$_2$. The parameter surveys of the TRAPPIST-1 planets performed in \citet{Turbet2018} provides a good starting point, although based on our results we recommend that future work focused on the specific role that background gases play in the habitable zone in general should also systematically vary instellation and stellar effective temperature.

\subsection{Additional Tests}\label{sec:tests}

\begin{table}
\centering
\begin{tabular}{p{1.5in}||p{3.5in}}
    \textbf{Experiment} & \textbf{Description} \\
    \hline
    No Rayleigh Scattering & Disabled Rayleigh scattering component of radiation scheme\\ \hline
    No Advection & Disabled horizontal advection in PlaSim's dynamical core \\ \hline
    No Sea-Ice Feedback & Set sea ice albedo and snow albedo equal to ocean albedo \\ \hline
    No Clouds & Disabled cloud formation and radiative feedback \\ \hline
    No Ozone & Disabled prescribed stratospheric heating from ozone \\ \hline
    No Oceans; Wet Soil & Uniform land surface initialized with saturated soil water \\ \hline
    No Water Vapor & Evaporation disabled; model initialized with zero atmospheric water \\ \hline
    No Water Vapor; No Advection & Evaporation disabled and horizontal advection disabled (dry, static column)\\ \hline
    No Water Vapor; No Rayleigh Scattering & Evaporation disabled and Rayleigh scattering disabled (transparent atmosphere)\\ \hline
    No Advection; No Water Vapor; No Scattering & Evaporation, Rayleigh scattering, and horizontal advection disabled (dry, static, transparent column) \\ \hline
    No Pressure Broadening & Removed surface pressure dependence from absorptivity pressure-scaling\\ \hline
    No Advection; No H$_2$O; No Scattering; No Pressure Broadening & All direct impacts from pN$_2$ on radiation or circulation removed; only vertical distribution of CO$_2$ can have an effect \\ \hline
    No Advection; No H$_2$O; No Scattering; No CO$_2$ & Fully transparent, static column; no pN$_2$ dependence possible \\ \hline
    10 mbar CO$_2$ vs Earth CO$_2$ & Sensitivity test to determine importance of pressure broadening of CO$_2$ vs H$_2$O; Earth-like land and obliquity, 0.1--10 bars, 1350 W/m$^2$ \\ \hline
    Snowball & Using cold-start initial conditions to probe sensitivity with little water vapor and a reflective surface
\end{tabular}
\caption{Description of the various experiments used to investigate the physical mechanisms underlying the trends observed in \autoref{fig:pn2grid}. In each case, we used PlaSim in a zero-obliquity aquaplanet configuration with warm-start initial conditions (unless otherwise specified) and considered a range of N$_2$ partial pressures ranging from 1--10 bars, and ran the models until they converged following the same procedure as described in \autoref{sec:strategy}.}\label{table:expdesc}
\end{table}

In order to identify the underlying physical mechanisms responsible for the nonlinear trends we observe in our model samples, we also ran a number of additional tests using subsets of the original model sample. These experiments are summarized in \autoref{table:expdesc}. In each case we used an aquaplanet configuration with warm-start initial conditions and zero obliquity and ran until energy balance equilibrium at the surface and top of the atmosphere were reached. In one case however instead of an aquaplanet configuration, we replaced the ocean with a uniform land surface and 5-meter-deep soil water reservoirs that were initialized fully-saturated. PlaSim uses a bucket model for soil water with horizontal advection of excess water (forming a river system), and includes exchange between the soil bucket and atmosphere through precipitation and evaporation. In another case, rather than using warm-start initial conditions, we used cold-start initial conditions to probe the sensitivity to pN$_2$ of snowball climates, where the surface is entirely covered by sea ice \citep{pierrehumbert05}. Due to the high albedo of sea ice on planets orbiting Sun-like stars, this state is bistable with temperate, mostly ice-free climates across a range of instellations and CO$_2$ levels \citep{Budyko1969,Sellers1969}, and features a relatively dry atmosphere\citep{pierrehumbert05}. We achieved our cold-start initial conditions by starting from the default warm initial conditions, reducing the instellation by 25 Wm$^2$ each year until the model had been fully-frozen for 30 consecutive years, then increased the instellation by 25 W/m$^2$ each year until our prescribed instellation of 1300 W/m$^2$ was reached. Once that point had been reached, we ran the model to equilibrium for a minimum of an additional 50 years. 

In many of the other experiments, rather than changing initial conditions or boundary conditions, we disabled some of the model physics or removed the surface pressure dependence from that physical process. For example, pressure broadening cannot be removed entirely because it affects how absorptivity and emissivity change between the top and bottom of the atmosphere. However, the reference pressure used in the pressure broadening computation can be set to the surface pressure rather than a fixed value, so that the amount of pressure broadening at the surface relative to the top of the atmosphere is constant across the parameter space. This does not change the reference pressure used in other parts of the radiation model or in the rest of the GCM. In experiments in which water vapor was removed, we increased the effective solar constant to offset the loss of the water vapor greenhouse, assuming a surface temperature of 280 K for a 1 bar atmosphere, such that $T_\text{eff}=\left[I_0(1-\alpha)/(4\sigma)\right]^{1/4}$, where $I_0$ is the instellation, $\alpha$ is albedo, and $\sigma$ is the Stefan-Boltzmann constant. In all other experiments, unless otherwise specified, we used an instellation fo 1300 W/m$^2$. 

\subsection{Reflected-light Spectra with SBDART}\label{sec:obsmethods}

We used the radiative transfer package SBDART \citep{Ricchiazzi1998} to compute reflectance and emission spectra for our models. Spatially-resolved images of exoplanets are not currently possible, so to compute the disk-averaged spectrum of each model, we run SBDART on each column of the model's output that would be visible to an observer for a given snapshot in time, using PlaSim's output to specify water abundances, cloud fractions, surface type, and air temperature, and specifying the solar zenith angle and observer's viewing angle according to the latitude and longitude of each model column. We assume the observer and Sun are both in the same place in the sky over the planet (corresponding to a planet as seen in secondary eclipse). We use the MODTRAN-3 input solar spectrum, which has a resolution of 20 cm$^{-1}$. We compute a single disk-averaged spectrum for each model by combining the spectrum for each individual column according to its surface area projected onto the line-of-sight: 
\begin{linenomath*}
\begin{equation}
F_\lambda(\Phi_{\text{obs}}) = \frac{\sum\limits_{\theta,\phi} F_\lambda(\theta,\phi,\Phi_{\text{obs}})A(\theta,\phi)\cos\Phi_{\text{obs}}}{\sum\limits_{\theta,\phi}A(\theta,\phi)\cos\Phi_{\text{obs}}}
\end{equation}
\end{linenomath*}
where $F_\lambda(\Phi_{\text{obs}})$ is the weighted average flux at each wavelength, $F_\lambda(\theta,\phi,\Phi_{\text{obs}})$ is the flux at a given latitude $\theta$ and longitude $\phi$, $A(\theta,\phi)$ is the area of that grid cell, and $\Phi_{\text{obs}}$ is the zenith angle of the observer (and therefore also the solar zenith angle).

Because Rayleigh scattering is responsible for the sky's distinctive blue color, the effect of increased Rayleigh scattering from higher pN$_2$ should be apparent in broadband photometry. We therefore integrate the resulting high-resolution spectra across B and V bands, $445\pm47$ and $551\pm44$ nm \citep{Binney1998}. We ignore any potential effects from astronomical detectors and Earth's atmospheric transmission windows, showing only the integrated flux from the planet in the noted wavelength ranges.

\section{Results}\label{sec:results}
\subsection{Earth-like Rotators}\label{sec:earthrot}
\begin{figure}
\begin{center}
\includegraphics[width=6in]{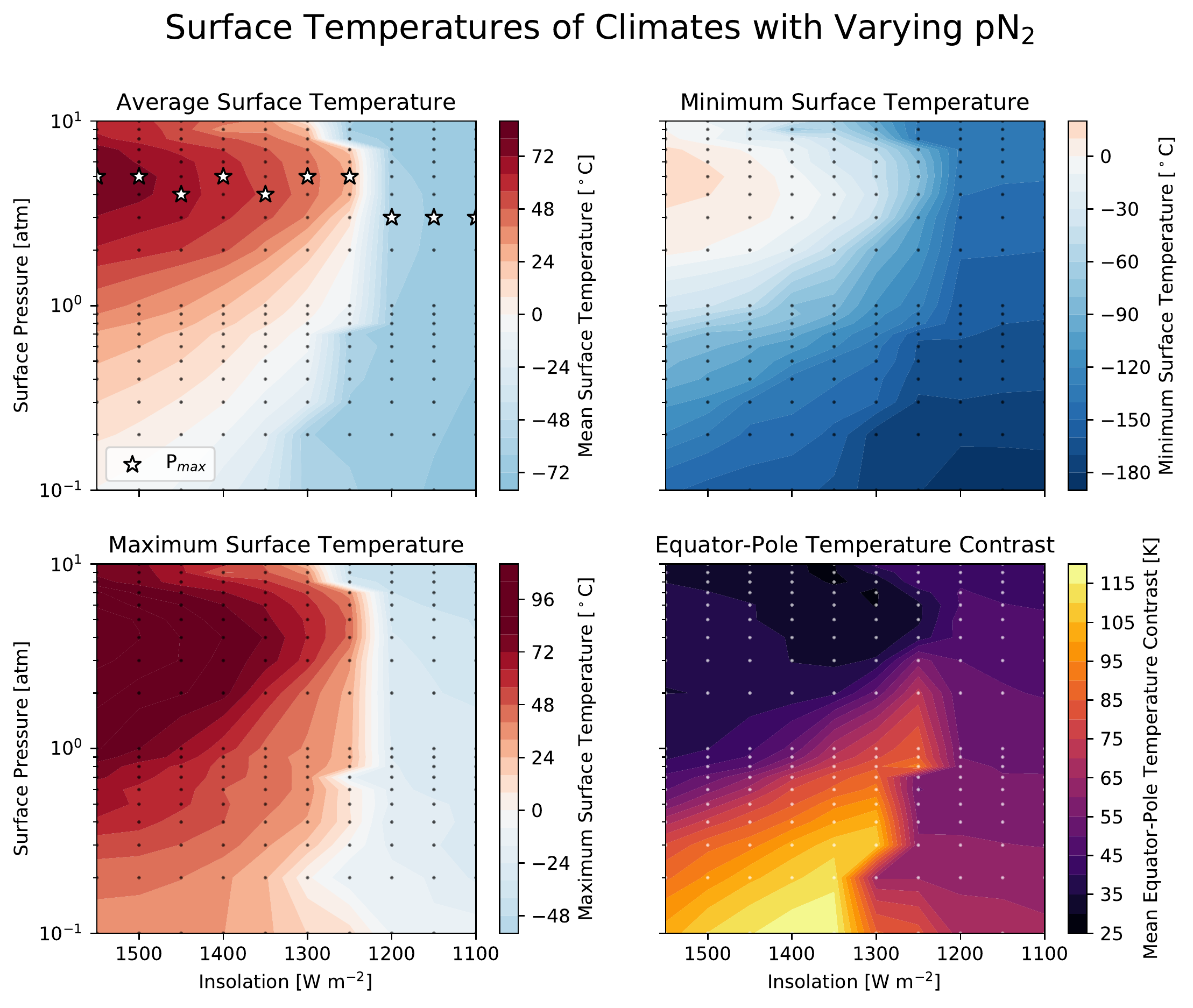}
\end{center}
\caption{Mean, minimum, and maximum surface temperatures, along with equator-pole surface temperature contrast, for a grid of 190 PlaSim models of Earth-like planets with varying surface pressures. Dots indicate the locations of individual models in the parameters space. In varying surface pressure, we only varied the Nitrogen partial pressure (pN$_2$)---pCO$_2$ was held constant. The climate demonstrates nonlinear sensitivity to the amount of background gas, showing significant warming in some regimes and significant cooling in others. This is due to warming by pressure broadening, cooling by Rayleigh scattering, and cooling by heat transport all competing with each other. The stars labeled P$_{max}$ in the upper-left panel indicate the surface pressures at which the mean temperature is maximized. We note that while very warm temperatures are reached in models at high instellations and high pressures, PlaSim lacks the physics necessary to model the transition to a runaway greenhouse, and we therefore cannot make conclusions about the inner edge of the habitable zone.}\label{fig:pn2grid}
\end{figure}

\begin{figure}
\begin{center}
\includegraphics[width=6in]{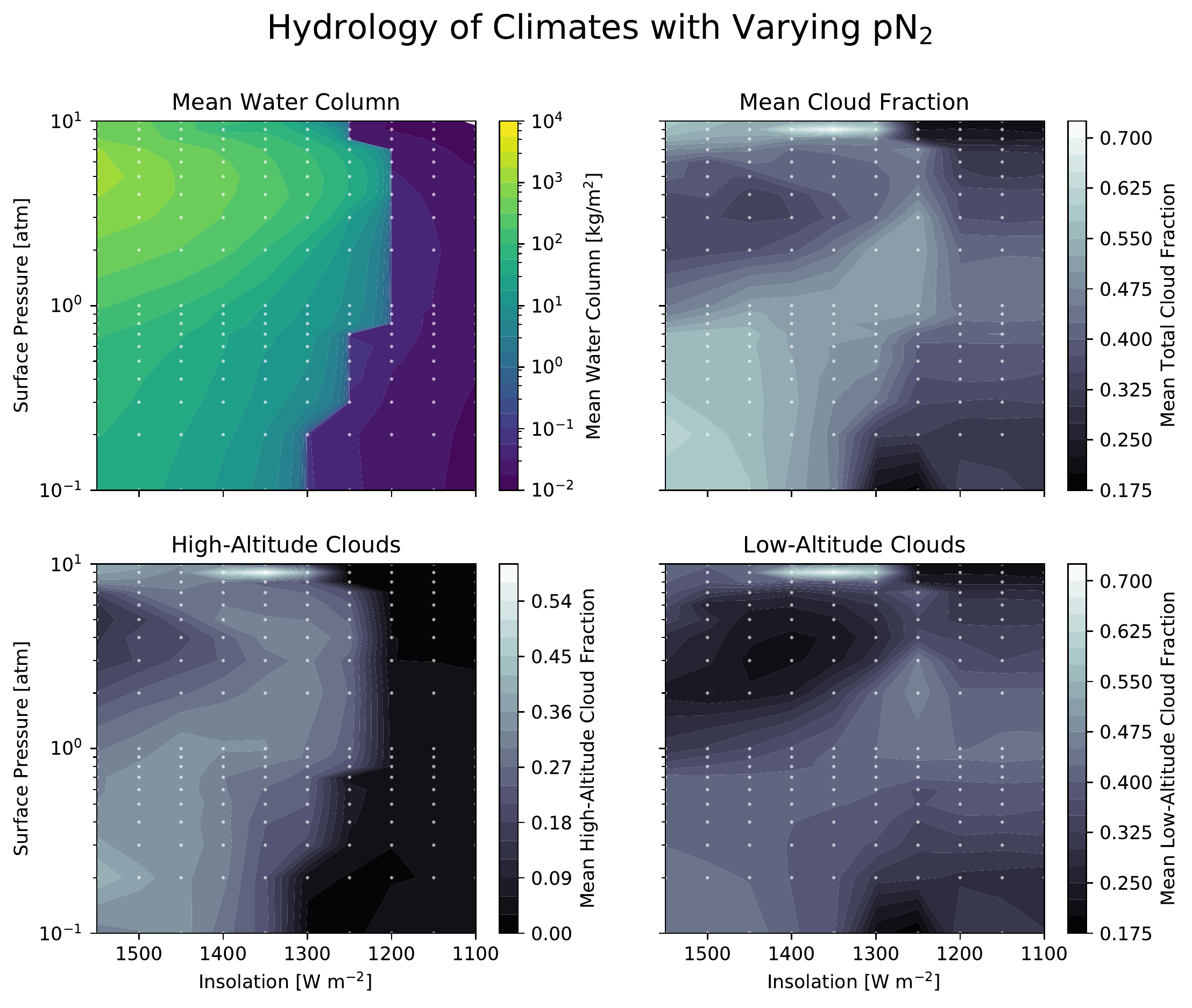}
\end{center}
\caption{Mean water column and total, low, and high cloud fraction for the same grid of models as in \autoref{fig:pn2grid}. Dots indicate the locations of individual models in the parameter space. Water vapor is correlated with mean temperature, but the cloud response is nonlinear and varies with instellation, appearing anti-correlated with temperature at high instellations and correlated with temperature at lower instellations, and therefore unlikely to be a driving cause of the climate's sensitivity to the background gas pressure. While large water column masses are produced at high pressures and high instellations, these water abundances nonetheless represent only a few percent of the total atmospheric column mass, such that water remains a minor component.}\label{fig:pn2grid2}
\end{figure}

As shown in \autoref{fig:pn2grid}, we find that the effect of increasing pN$_2$ on global temperatures is nonlinear: at low pressures, increasing pN$_2$ results in warmer average temperatures, while at higher pressures, increased pN$_2$ can cool the climate, resulting eventually in some cases in a transition to a snowball state, where sea ice extends all the way to the equator \citep{pierrehumbert05}. The warming associated with pN$_2$ is significant; planets with 4 or 5 bars of N$_2$ can be an average of almost 40 K warmer than planets with 1 bar. For planets in snowball states, we do not observe a significant warming trend, as shown in low-instellation models in \autoref{fig:pn2grid}. We further find that the N$_2$ partial pressure at which average surface temperatures reach their maximum appears weakly dependent on instellation, such that planets at high instellations begin net cooling at slightly higher surface pressures than planets at lower instellations. Maximum and mean surface temperatures reach very warm levels at high instellations and pressures, which might lead the reader to suspect that some of our models should transition to runaway greenhouse states. However, while PlaSim has been used before to study this transition \citep{Gomez-Leal2018}, it cannot model significant changes in pressure due to rapid evaporation, assumes that water is a minor constituent of the atmosphere, and its radiation scheme has not been validated for very warm, very moist atmospheres. In our models, as shown in \autoref{fig:pn2grid2}, the water column in the hottest models remains only a few percent of the total column mass, and so constitutes a minor constituent even in our warmest models. We avoid using the stratospheric humidity criteria used in \citet{Gomez-Leal2018}, as the efficiency of the tropopause water trap depends on surface pressure \citep{Zahnle2016}, making this metric unreliable at surface pressures different from Earth's. Previous studies of the inner edge of the habitable zone using GCMs have used the point at which the model breaks or is no longer able to achieve radiative equilibrium as a proxy for the inner edge \citep[e.g.][]{Leconte2013,Yang2014b,Bin2018}; due to PlaSim's simplicity and lack of relevant physics for steam atmospheres, we suspect this would also be an unreliable metric. We therefore refrain from concluding that these models represent the inner edge of the habitable zone. 

% We examined the possibility of a model-dependent source for this behavior by running the LMD Generic model for an aquaplanet at different pN$_2$ levels. We observed a warming trend in those models as well, although we did not observe a subsequent cooling trend at higher pressures, though the highest pressure we considered there was 7 bars. It is possible that LMD Generic would recover the cooling trend at even higher pressures. We also note that a similar nonlinear sensitivity to pN$_2$ was observed in ExoCAM models reported in \citet{Komacek2019}, but their results included a limited number of models and had no CO$_2$, making it difficult to draw conclusions about the mechanisms responsible for this trend. All three of these GCMs find different pressures for the maximum average temperature, which could indicate variability in how the mechanisms responsible are modeled, but may also be due to differences in model setup, such as the exclusion of CO$_2$ in the \citet{Komacek2019} models. 

We note that non-monotonic trends in surface temperature with varying pN$_2$ were also observed in ExoCAM models reported in \citet{Komacek2019}, but their results included a limited number of models and had no CO$_2$, making it difficult to draw conclusions about the mechanisms responsible. Similar behavior was also observed with 1D models in \citet{Goldblatt2009} and \citet{Keles2018}. The fact that the same qualitative behavior appears in all four models however suggests that this is not model-dependent behavior, and is instead representative of underlying physics.

\begin{figure}
\begin{center}
\includegraphics[width=6in]{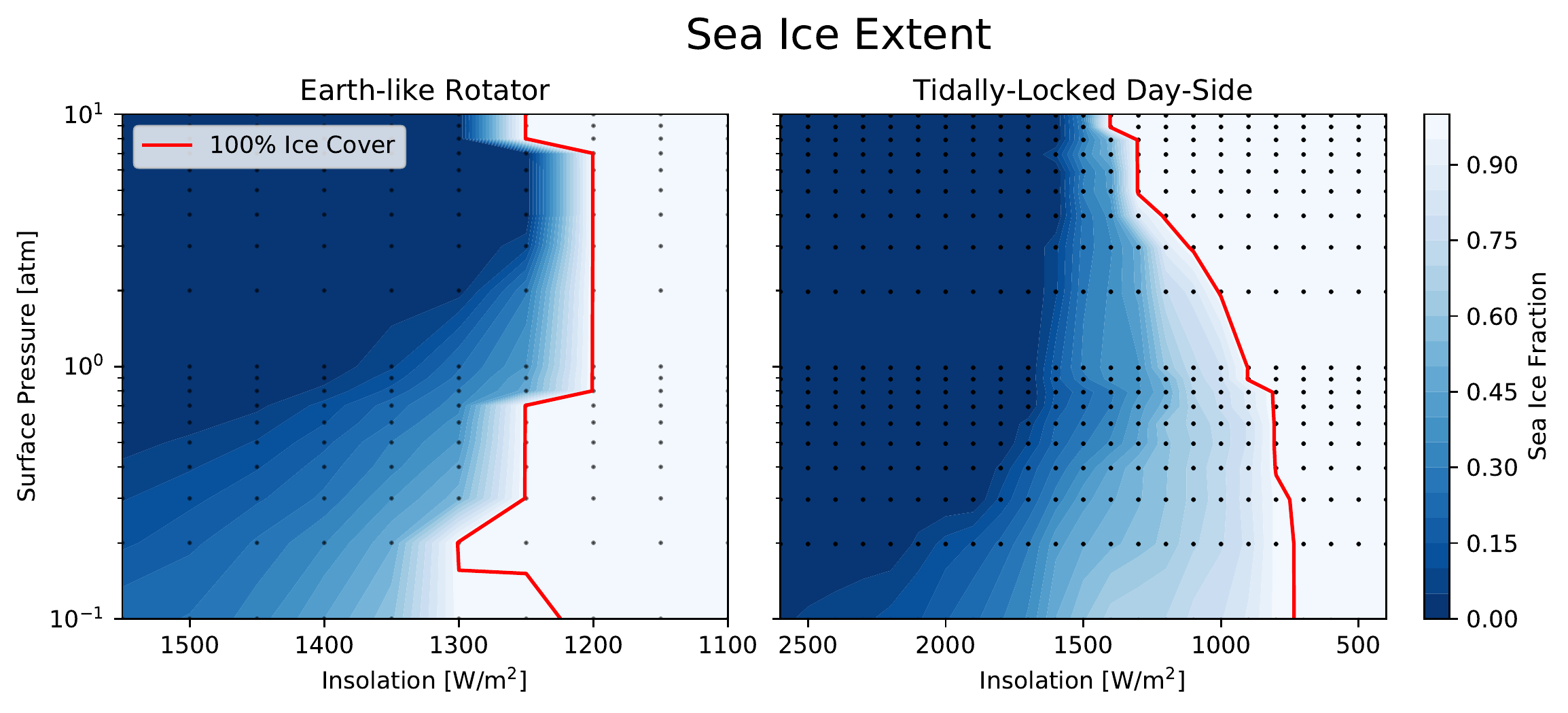}
\end{center}
\caption{Sea ice fraction for both the Earth-like models shown in \autoref{fig:pn2grid} and the tidally-locked models shown in \autoref{fig:lockedgrid}. Dots indicate the locations of individual models in the parameters space. For the latter, we show only the dayside sea ice fraction, as the sea ice albedo feedback does not operate on the night side of tidally-locked planets. In both sets of climates, we find that the transition from fully-frozen to ice-free is much sharper at higher pressures, likely due to weaker equator-pole temperature contrasts.}\label{fig:seaice}
\end{figure}

\begin{figure}
\begin{center}
\includegraphics[width=6in]{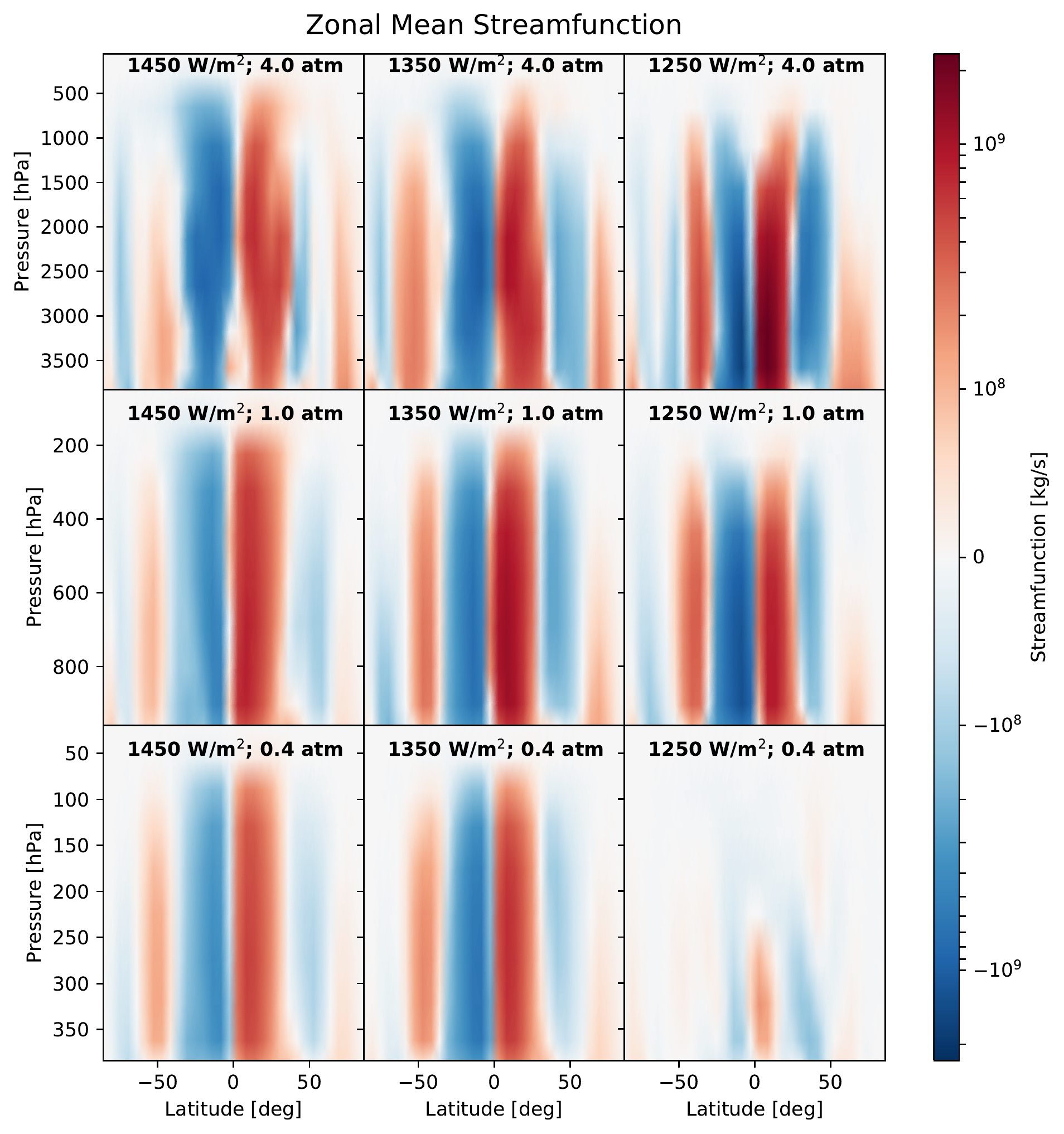}
\end{center}
\caption{Zonally-averaged meridional streamfunction as a function of pressure level for 9 models selected from the grid shown in \autoref{fig:pn2grid}. Here positive values indicate clockwise rotation. We find that poleward heat transport is generally more efficient at higher pressures. The bottom-right panel is a model in a snowball state, and has generally smaller and weaker meridional circulation, with an enhanced ferrel cell due to elevated temperatures over the northern hemisphere landmasses.}\label{fig:2dstf}
\end{figure}

\begin{figure}
\begin{center}
\includegraphics[width=6in]{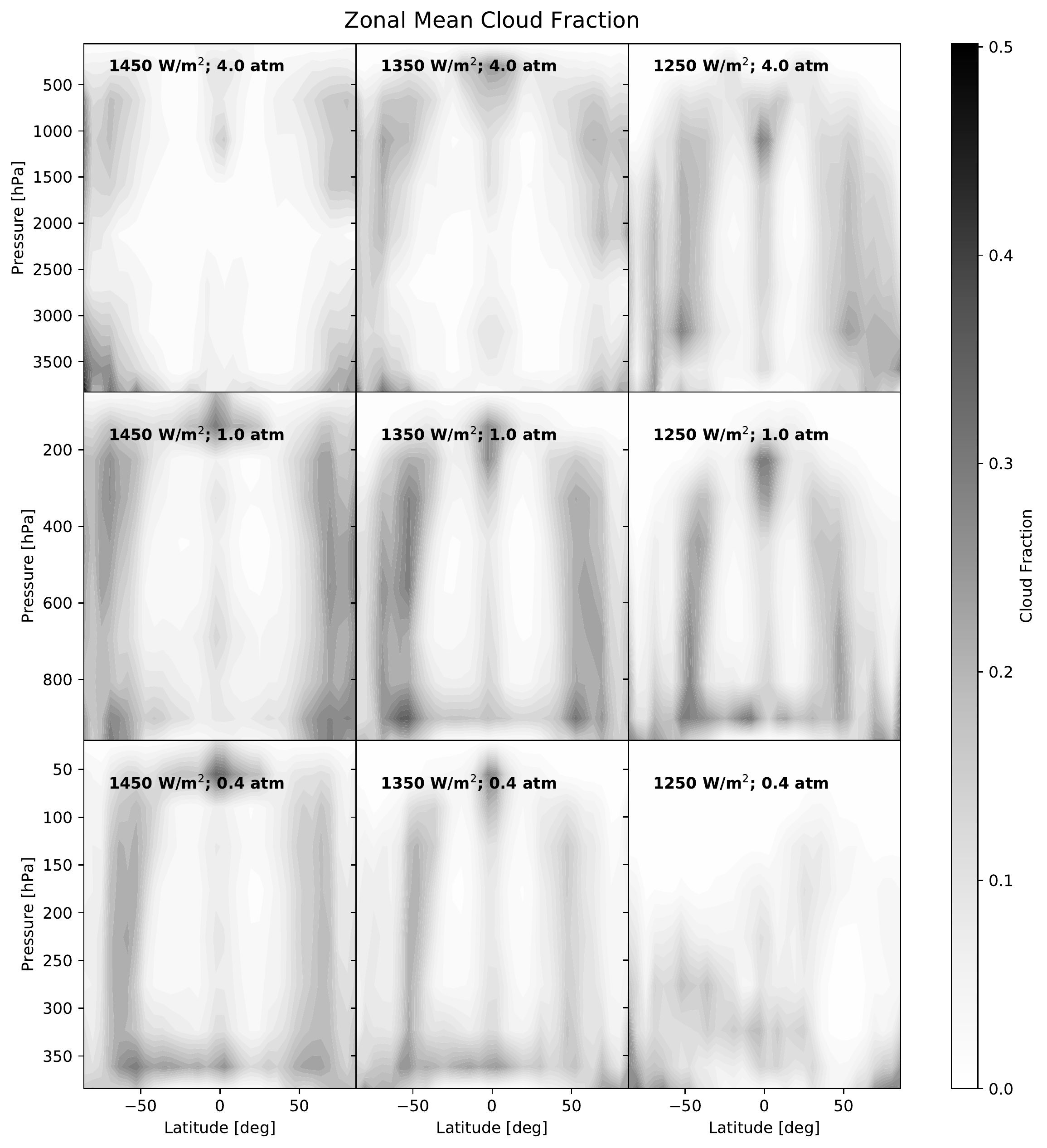}
\end{center}
\caption{Zonally-averaged clouds for 9 models selected from the grid shown in \autoref{fig:pn2grid}. Mid-latitude cloud belts appear to move to higher latitudes at higher pressures. The bottom-right panel is in a snowball state.}\label{fig:2dclouds}
\end{figure}

Atmospheric water vapor and clouds are shown in \autoref{fig:pn2grid2}, and sea ice fraction for both these models and the tidally-locked models presented in \autoref{sec:tlresults} is shown in \autoref{fig:seaice}. The cloud response is nonlinear, and its relationship with the mean annual surface temperature appears to vary with instellation, such that cloud fraction is anti-correlated with temperature at high instellations and very high temperatures, and positively correlated with surface temperature at lower instellations. Cloud distribution in our models is primarily localized near the equator and at the upwelling branches of the Ferrell cells, as can be seen by comparing \autoref{fig:2dstf} and \autoref{fig:2dclouds}. We therefore conclude that the cloud response is unlikely to drive the overall trends we observe in surface temperature. We further note that cloud fraction seems to decrease most in the models that achieve the highest temperatures, so we caution the reader that our earlier warnings about interpreting our hottest models should apply here as well. 

Atmospheric water vapor is correlated with mean temperature, such that our warm models are also moist. This is to be expected, as evaporation rates in PlaSim are driven by the Clausius-Clapeyron relation \citep{Fraedrich2005}, though the total column mass fraction of water vapor is no more than a few percent in our models, reaching a maximum of 7\% in the 0.1 bar model at 1550 W/m$^2$. We note however that the transition from ice-free to fully-frozen climates as instellation decreases is much steeper at high pressures. A possible explanation for this observation is that in the limit of very efficient heat transport, as is found at high pressures, surface temperature gradients are small. A small horizontal temperature gradient means sea ice extent is more sensitive to the average temperature, so the transition from ice-free to snowball is sharper.

\begin{figure}
\begin{center}
\includegraphics[width=6in]{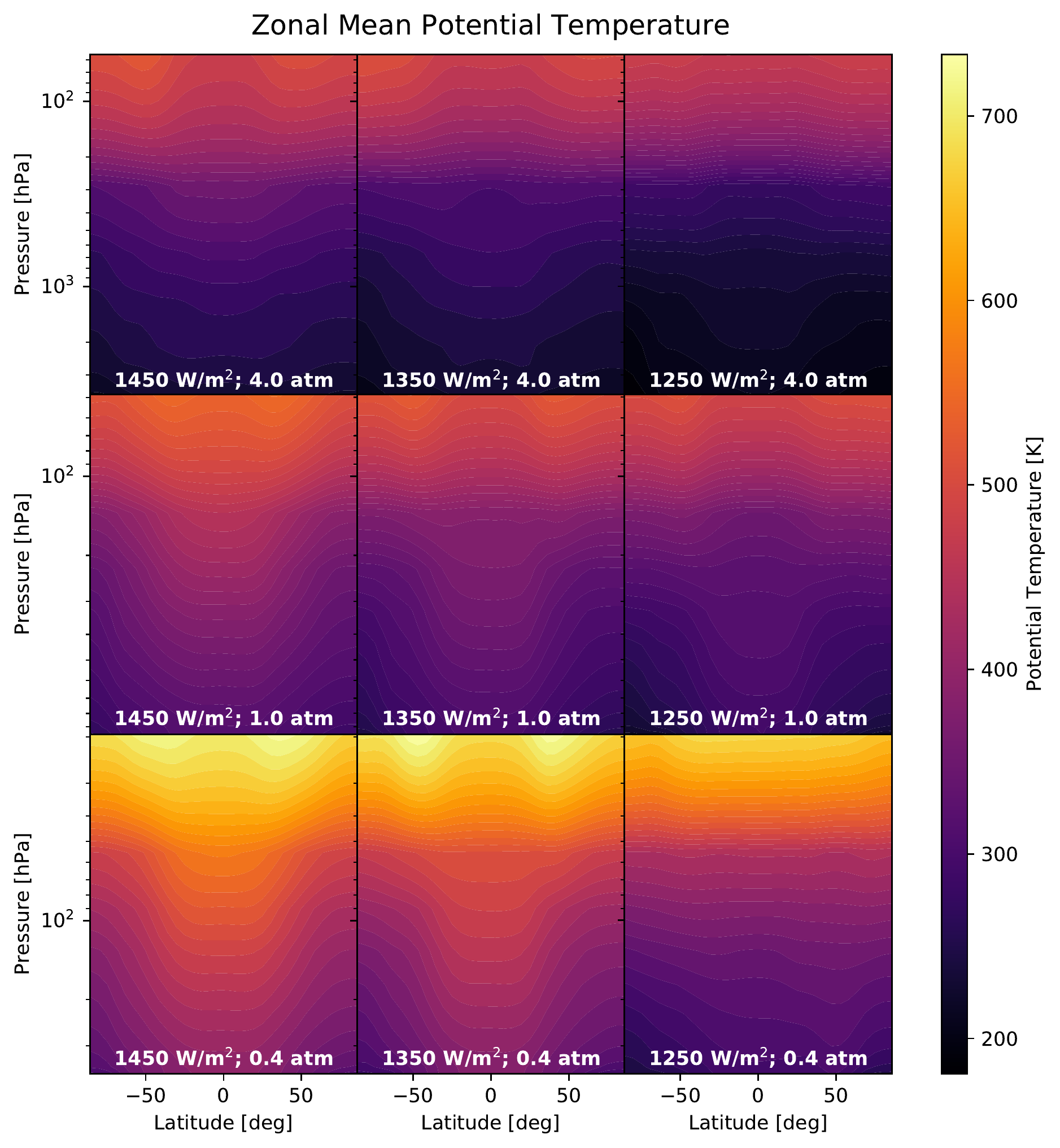}
\end{center}
\caption{Zonally-averaged potential temperature as a function of pressure level for 9 models selected from the grid shown in \autoref{fig:pn2grid}. The potential temperature gradient at the tropopause is steeper at higher pressures, consistent with the stronger water trap discussed in \citet{Zahnle2016} and the weaker vertical heat transport found in \citet{Kaspi2015}. The snowball climate in the bottom-right panel is relatively stable against convection, and has a slightly warmer northern hemisphere, caused by a lower average albedo due to areas of bare land.}\label{fig:2dtemp}
\end{figure}

\begin{figure}
\begin{center}
\includegraphics[width=6in]{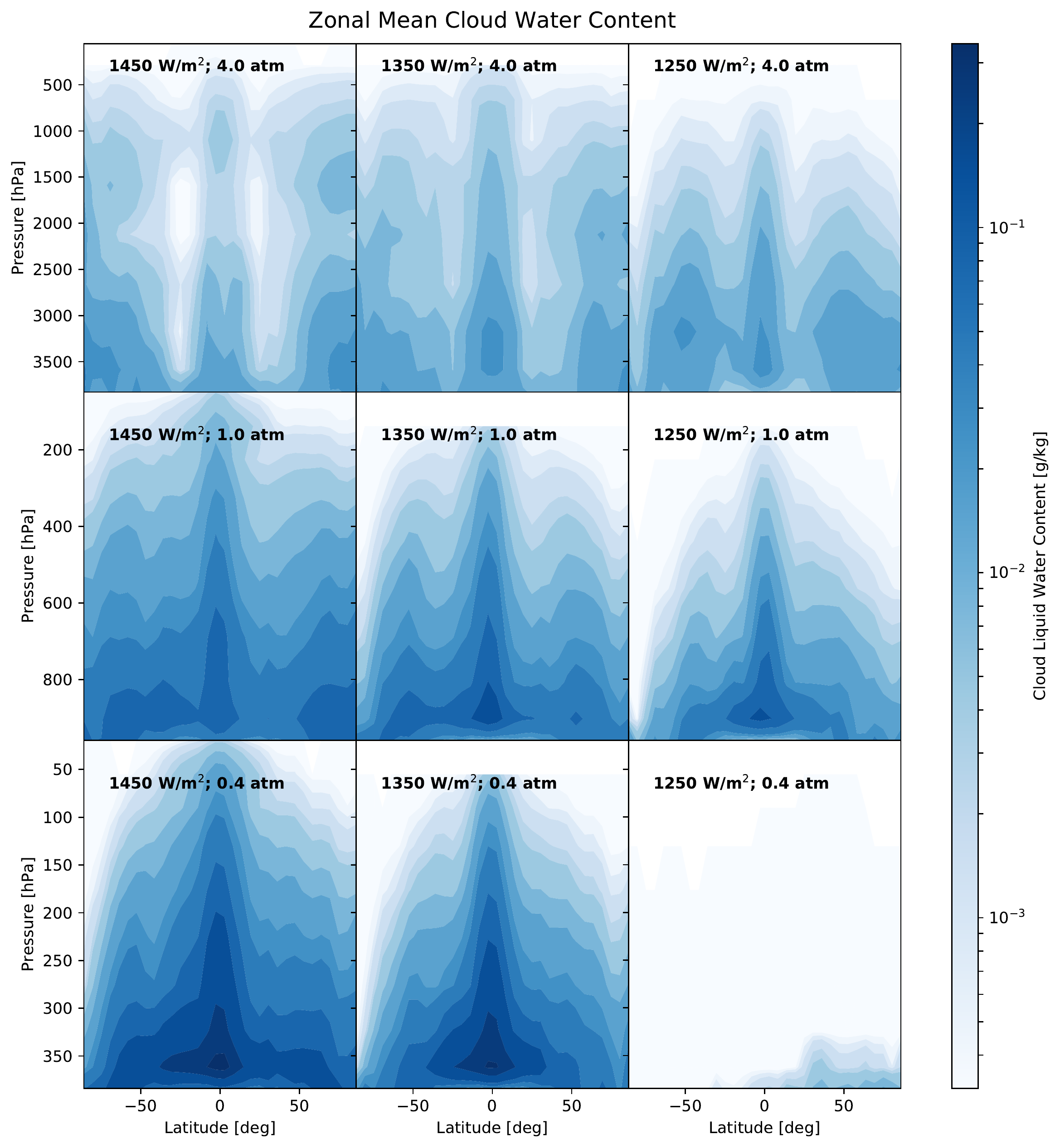}
\end{center}
\caption{Zonally-averaged Cloud liquid water content as a function of pressure level for 9 models selected from the grid shown in \autoref{fig:pn2grid}. While our high-pressure climates are warmer and therefore wetter, the air in clouds is drier, and moist convection in the tropics is suppressed. The snowball climate in the bottom-right panel has a very dry atmosphere, and thus very little cloud water.}\label{fig:2dwater}
\end{figure}

A more detailed look at some of the climates in our sample is given in \autoref{fig:2dtemp}, \autoref{fig:2dstf}, \autoref{fig:2dclouds}, and \autoref{fig:2dwater}. We find that our high-pressure climates are characterized by a slight poleward shift of the mid-latitude cloud belts, steeper potential temperature gradients at the tropopause, more-efficient poleward heat transport, and drier clouds. The lower cloud water specific humidity shown in \autoref{fig:2dwater} is partly related to the greater stability against convection suggested by the potential temperature gradients in \autoref{fig:2dtemp}---less-vigorous convection draws less water into the upper parts of the troposphere. We also find however that in the lower atmosphere, relative humidity is not particularly sensitive to surface pressure, suggesting that most of the overall decrease in cloud water specific humidity is not due to a proportional decrease in atmospheric water, but rather a consequence of the fact that the water saturation vapor pressure is a function only of temperature, so at higher pN$_2$, a given parcel of air will reach saturation at a lower specific humidity. One of the snowball climates in our sample is shown in the bottom-right panel of each figure. Our snowball models have dry atmospheres and significantly weaker meridional circulation, consistent with the findings in \citet{pierrehumbert05}. Our models have increased temperatures at northern hemisphere mid-latitudes and therefore North-South asymmetries in streamfunction, cloud field, and atmospheric water content. This is due to land areas without snow and therefore lower albedo, as discussed in \citet{Paradise2019}.

\subsection{Additional Experiments}\label{sec:expresults}

\begin{table}
\centering
\begin{tabular}{c||c|c}
    Experiment & Warming Trends & Cooling Trends \\
    \hline
    No Rayleigh Scattering & Yes & \textbf{No} \\
    No Advection & Yes & Yes \\
    No Sea-Ice Feedback & Yes & Yes \\
    No Clouds & Yes & Yes \\
    No Ozone & Yes & Yes \\
    No Oceans; Wet Soil & Yes & Yes \\
    No Water Vapor & \textbf{No} & Yes \\
    No Water Vapor; No Advection & \textbf{No} & Yes \\
    No Water Vapor; No Rayleigh Scattering & \textbf{No} & Yes \\
    No Advection; No Water Vapor; No Scattering & Yes & \textbf{No} \\
    No Pressure Broadening & \textbf{No} & Yes \\
    No Advection; No H$_2$O; No Scattering; No Broadening & \textbf{No} & Yes \\
    No Advection; No H$_2$O; No Scattering; No CO$_2$ & \textbf{No} & \textbf{No}
\end{tabular}
\caption{Summary of the results of the experiments described in \autoref{table:expdesc}. For each experiment, we identify whether increasing pN$_2$ resulted in an average increase in surface temperature, an average decrease, or both (as seen in the nonlinear response in \autoref{fig:pn2grid}). Results are bolded where the outcome differs from the behavior found in \autoref{fig:pn2grid}. These results suggest that the dominant cooling mechanism for Earth-like rotators is increased Rayleigh scattering, and the dominant warming mechanism is pressure broadening of CO$_2$ and H$_2$O absorption. Without water vapor, pressure-broadening of CO$_2$ does not result in enough additional forcing to counteract cooling by scattering and heat transport. Increased heat transport cools the climate, but in these experiments was insufficient to counter water-amplified warming alone, in contrast to Rayleigh scattering, which was sufficient.}\label{table:experiments}
\end{table}

To investigate the physical mechanisms underlying the trends observed in \autoref{fig:pn2grid}, we conducted several experiments changing initial conditions, boundary conditions, or model physics, described in \autoref{sec:tests}. Their results, excluding the elevated CO$_2$ and snowball experiments, are summarized in \autoref{table:experiments}.

Without heat transport or Rayleigh scattering, the global average temperature increases monotonically with increasing pN$_2$, until our model crashes.  The effect of pressure broadening from CO$_2$ on its own is small---no more than 5--10 K between 1 and 10 bars of pN$_2$. If either cooling from increased Rayleigh scattering or increased heat transport is present, that alone is enough to counteract the warming from pressure broadening of CO$_2$, resulting in net cooling in dry atmospheres---pressure broadening of H$_2$O absorption and amplification from the water vapor positive feedback are necessary for pressure-broadening to be the dominant mechanism. In the absence of water vapor, heat transport, scattering, or pressure broadening, a slight cooling trend remains---approximately 0.5 K over 1--10 bars. This cooling trend disappears when CO$_2$ is removed as well, suggesting that the vertical distribution of CO$_2$ in taller atmospheres is very slightly more efficient at cooling the surface. 

\begin{figure}
\begin{center}
\includegraphics[width=6in]{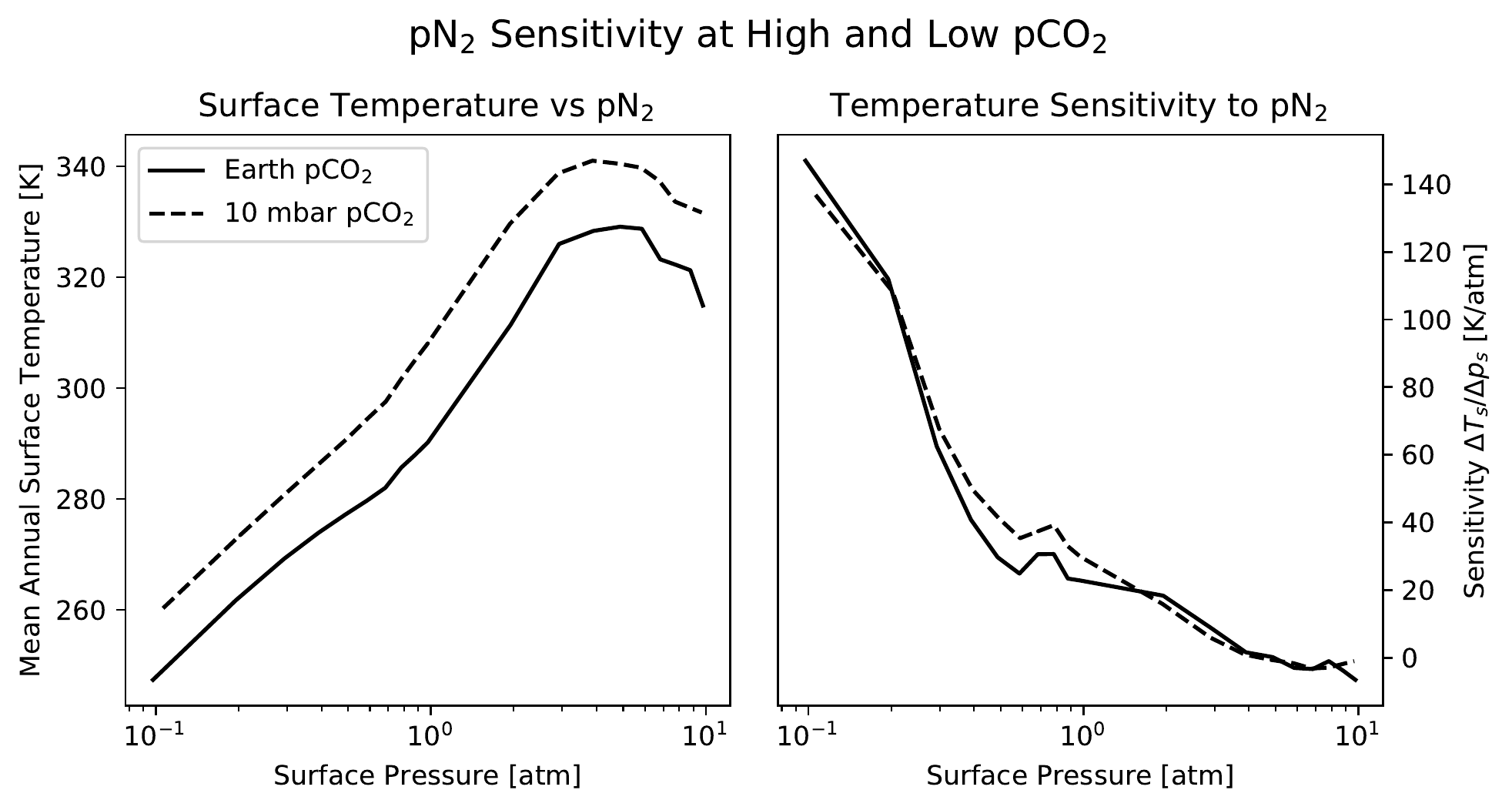}
\end{center}
\caption{Global mean annual surface temperature as a function of pN$_2$ (left) and the change in mean surface temperature per additional atm of N$_2$ (right), for Earth-like planets at 1350 W/m$^2$ with Earth-like pCO$_2$ (solid line) and with 10 mbar of CO$_2$ (dashed line). We find that despite increased forcing from CO$_2$ at 10 mbar, there is no significant change in sensitivity to pN$_2$, suggesting that pressure broadening of H$_2$O is much more climatically-relevant.}\label{fig:hico2}
\end{figure}

To explore whether the warming by pressure broadening is primarily caused by broadening or CO$_2$ absorption lines or H$_2$O absorption, we re-ran the 1350 W/m$^2$ slice of the parameter space in \autoref{fig:pn2grid} with 10 mbar of CO$_2$ instead of 360 $\mu$bar. The results are shown in \autoref{fig:hico2}. While the higher-CO$_2$ climates are significantly warmer, as expected, they demonstrate nearly identical sensitivity to increased pN$_2$ as those with Earth-like pCO$_2$. This suggests that absorption by water may be far more important for determining the climate's sensitivity. This is consistent with our finding that availability of water vapor, not CO$_2$, is necessary for an overall warming trend with increased pN$_2$, and is consistent with \citet{Komacek2019} finding the same nonlinear response that we find, despite omitting CO$_2$ from their models.  

\begin{figure}
\begin{center}
\includegraphics[width=5in]{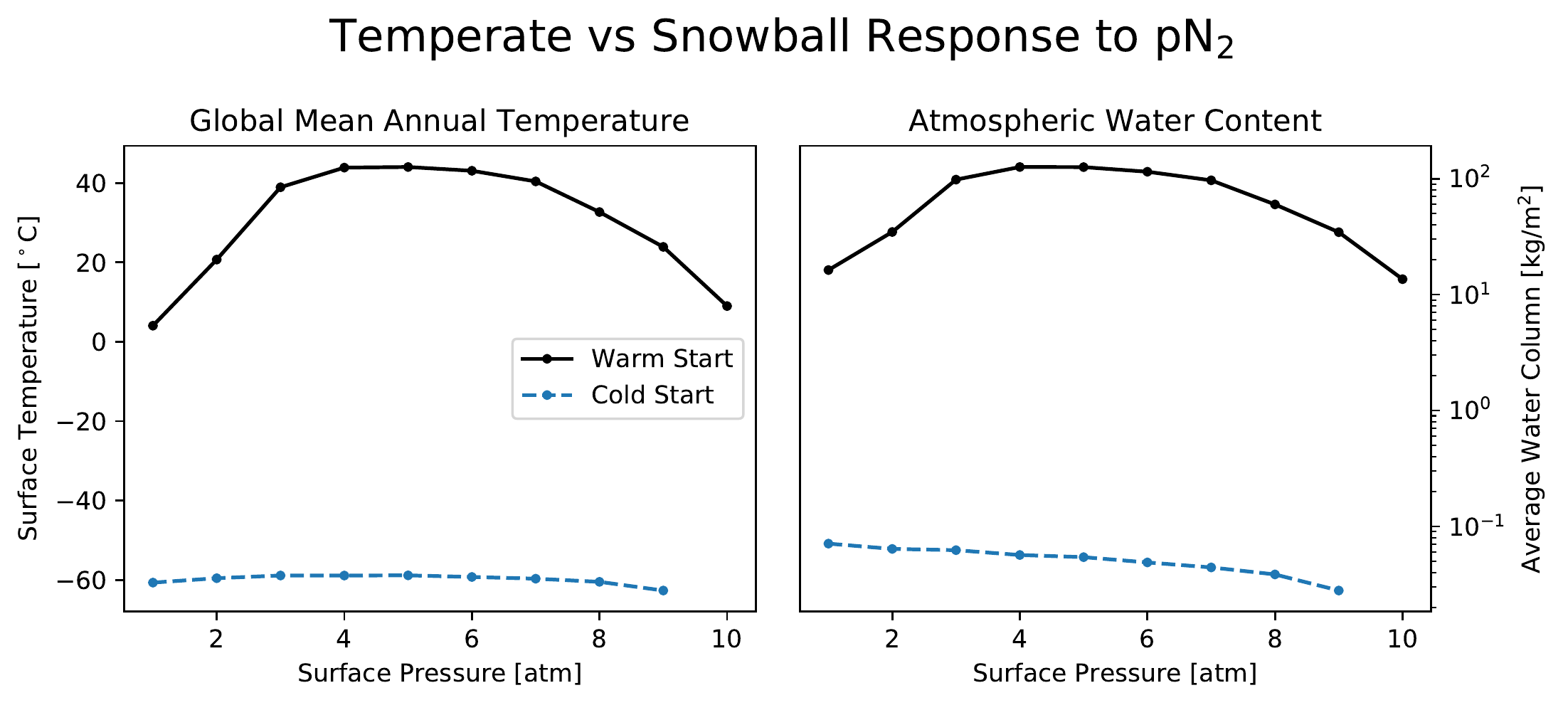}
\end{center}
\caption{Global mean annual surface temperature and average water column for varying surface pressures, at 1300 W/m$^2$, in both warm-start and cold-start (snowball) scenarios. In our cold-start scenarios, the ocean surface is entirely covered in sea ice. Warm-start models display the strongly nonlinear response identified in \autoref{fig:pn2grid}, while snowball models do not. The relatively large availability of water vapor in the temperate models means that pressure broadening of the CO$_2$ and water vapor absorption lines causes a large amount of warming, such that temperatures only decrease with a large amount of Rayleigh scattering at high surface pressures. On snowball planets, however, the limited availability of water vapor results in pressure broadening causing a small amount of warming at lower surface pressures, but Rayleigh scattering and heat transport quickly dominate, resulting in monotonically-decreasing water vapor abundance. In general, snowball planets are less sensitive to changes in pN$_2$, because the high surface albedo limits the degree to which atmospheric scattering can affect the overall albedo.}\label{fig:snowball}
\end{figure}

We examined the response of both temperate and snowball climates to differences in pN$_2$, as shown in \autoref{fig:snowball}. The temperate models display the strong, nonlinear response shown in \autoref{fig:pn2grid}, with increased heating through pressure broadening amplified by abundant atmospheric water vapor, and eventual cooling by Rayleigh scattering. On the other hand, the snowball models appear much less sensitive to pN$_2$, varying by only 4 K over 10 bars, despite the inclusion of the water vapor greenhouse effect.

\subsection{Synchronous Rotators}\label{sec:tlresults}

\begin{figure}
\begin{center}
\includegraphics[width=5in]{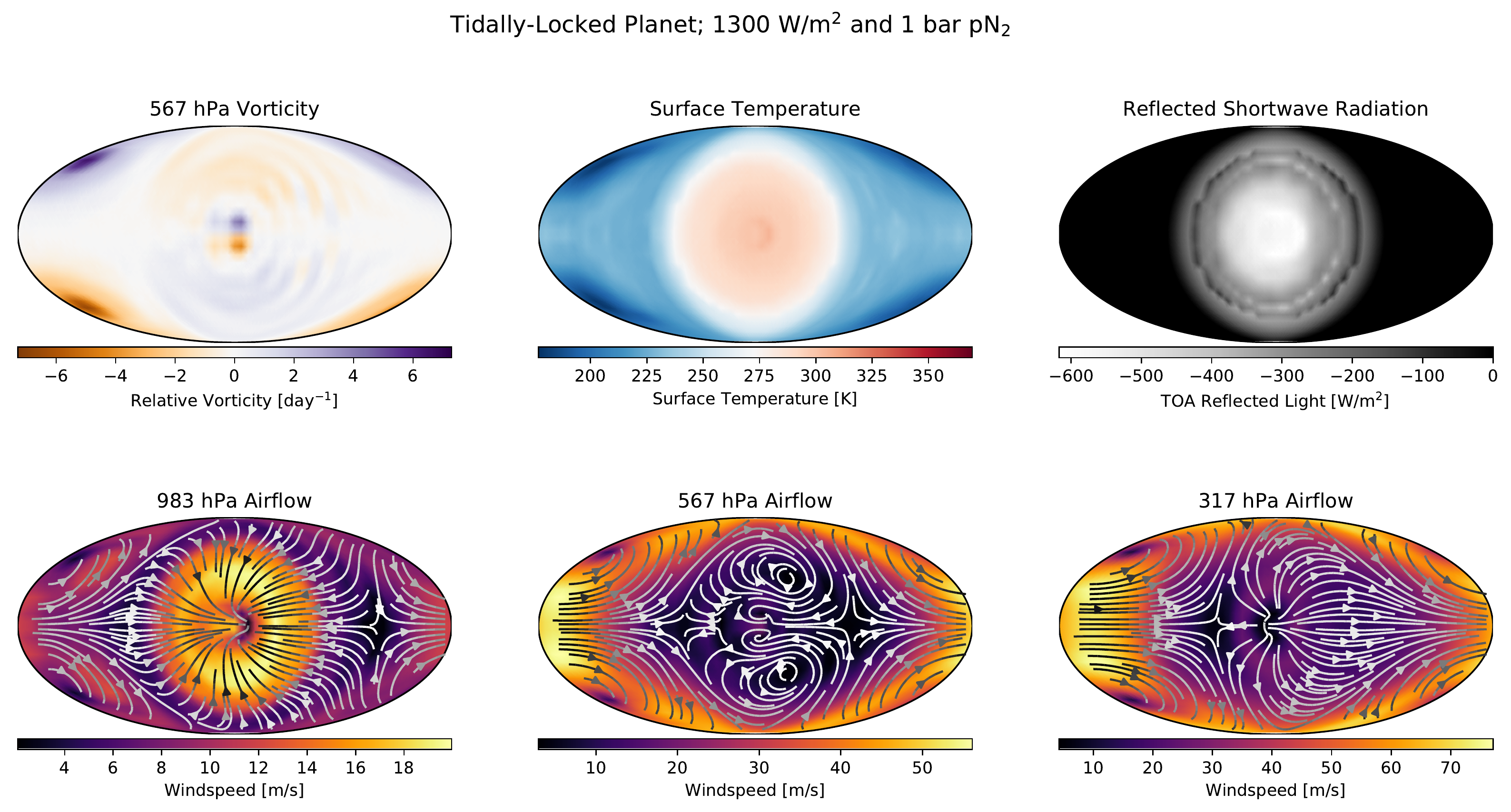}
\end{center}
\caption{Surface temperature, top-of-atmosphere reflected light, mid-atmosphere relative vorticity, and airflow at three different levels of the atmosphere, for one of our tidally locked models (1300 W/m$^2$ instellation and 1 bar of pN$_2$). The models are shown in a Mollweide projection, with the substellar point in the center. Our tidally-locked models show the same qualitative behavior as found in other models, including an eyeball-like region of open ocean \citep{Pierrehumbert2011}, clouds at the substellar point \citep{Edson2011,Yang2013}, night-side high-latitude gyres \citep{Edson2011}, and a substellar region characterized by surface-level inflow and net-eastward outlow in the upper troposphere \citep{Pierrehumbert2011,Edson2011}.}\label{fig:samplelock}
\end{figure}

We find that our tidally-locked models reproduce the major qualitative climate features observed in other simulations of tidally-locked planets, as shown in \autoref{fig:samplelock}. For temperate tidally-locked planets with Earth-like surface pressures, we find a nearly-circular area of above-freezing temperatures on the dayside, and nearly-uniform cold temperatures across much of the nightside \citep{Pierrehumbert2011}, with a slight eastward offset of the substellar hot spot. The global temperature contrast increases as instellation decreases, as found by \citet{Haqq-Misra2018}. Near the surface, air flows across the dayside toward the substellar point, where it rises, and then diverges in the upper troposphere, forming an eastward equatorial jet, guided by two strong high-latitude gyres on the nightside \citep{Pierrehumbert2011,Edson2011}. The dayside upwelling region is humid, and thus has extensive cloud cover \citep{Edson2011,Yang2013}. We have not examined the sensitivity of these climate features to high or low pN$_2$, and leave that investigation to a future study.

\begin{figure}
\begin{center}
\includegraphics[width=6in]{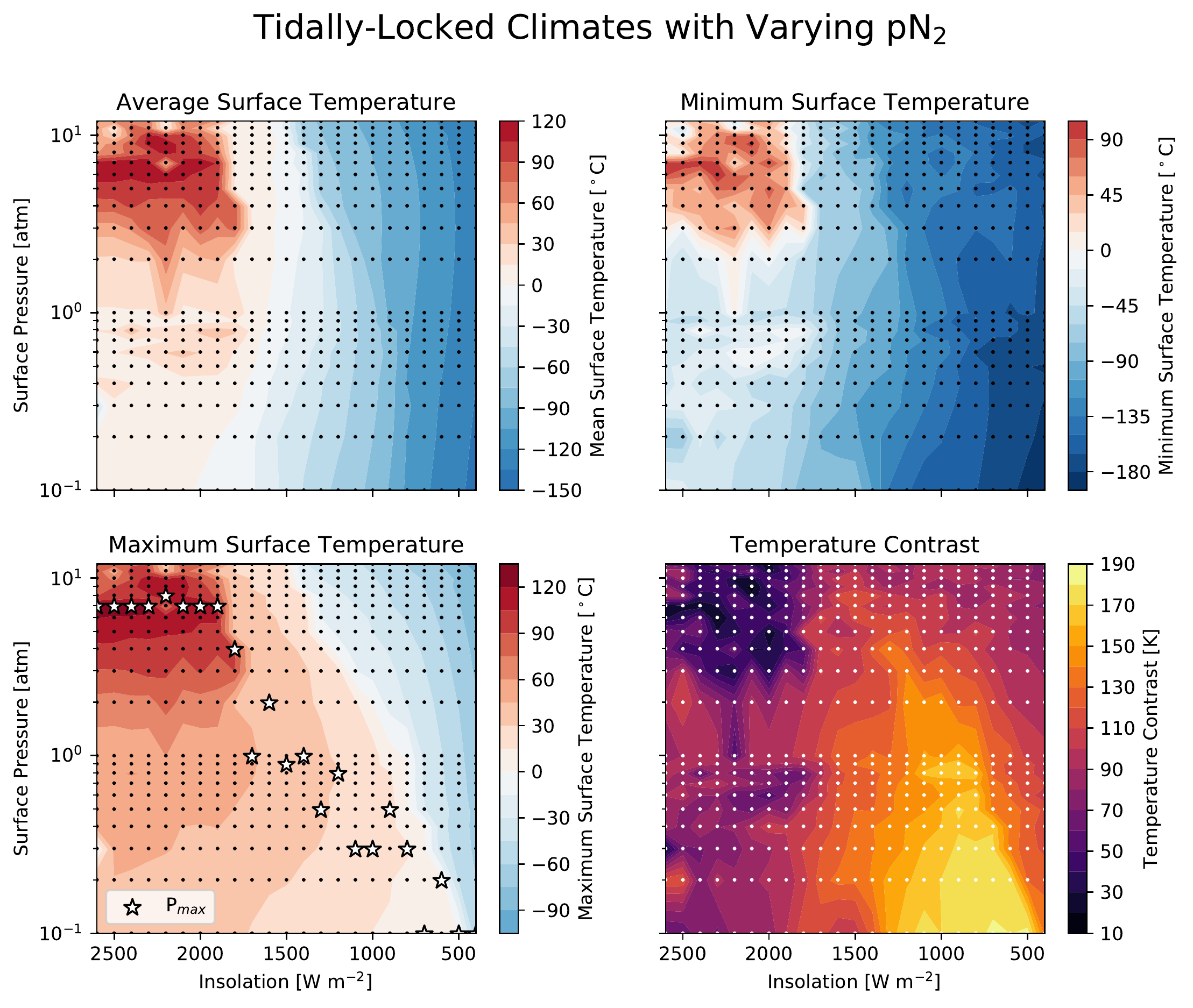}
\end{center}
\caption{Mean, minimum, and maximum surface temperatures, along with global temperature contrast, for a grid of 483 PlaSim models of tidally-locked planets with varying surface pressures. Dots indicate the locations of individual models in the parameter space. As in \autoref{fig:pn2grid}, the stars labeled P$_{max}$ in the bottom-left panel indicate the pressure at which the maximum temperature peaks. Only pN$_2$ was varied---pCO$_2$ was held constant. In contrast to \autoref{fig:pn2grid}, on tidally-locked planets pN$_2$ primarily cools the climate at lower and moderate instellations, while at higher instellations we see the same nonlinear behavior resulting from competing warming and cooling mechanisms as on Earth-like planets. This difference is because heat transport is much more important on tidally-locked planets, where advection of heat onto the night-side is a major component of the planet's total energy budget. There appears to be complex structure at high instellations, but as these climates are on the extreme end of what PlaSim can model, more work is required to determine its significance.}\label{fig:lockedgrid}
\end{figure}

The minimum, mean, and maximum temperatures of our tidally-locked models, along with the global temperature contrast, are shown in \autoref{fig:lockedgrid}. The day-side sea ice fraction is shown in \autoref{fig:seaice}. In this case we choose to show day-side sea ice instead of global sea ice because ice on the night-side does not reflect any light, and therefore the effect of varying background gas pressure on night-side ice is purely a consequence of ambient temperature, as opposed to ice on the day-side, where an ice albedo feedback exists in our models due to the solar-like input spectrum, and which may be affected by the Rayleigh scattering from elevated pN$_2$. There appears to be some amount of complex structure in night-side temperatures in our high-instellation, high-pressure models, but due to few constraints on night-side climate in these regimes, we cannot rule out a numerical origin and recommend further study. On these planets, the climate is very sensitive to the efficiency with which heat can be advected from the dayside to the night side \citep{Checlair2017,Haqq-Misra2018}. As pN$_2$ increases, heat transport efficiency increases, resulting in significant dayside cooling. Combined with increased Rayleigh scattering, warming by pressure broadening only dominates at low pN$_2$ where heat transport and scattering are weak, and at high instellations where there is high potential for surface evaporation, resulting in very strong amplification through the water vapor greenhouse. In contrast to the Earth-like models in \autoref{fig:pn2grid}, the surface pressure at which dayside surface temperatures are maximized has a strong dependence on surface pressure, moving to significantly lower pressures as instellation decreases, as shown in \autoref{fig:lockedgrid}. Here we are using changes in maximum temperature as the metric for the impact on the climate, in contrast to the average surface temperature, which we used in \autoref{sec:earthrot}. This is because the geometry of tidally-locked climates and the importance of horizontal heat transport on these planets means that changes in the global average surface temperature may not be particularly indicative of trends in the dayside climate. We find that just as with global sea ice extent on the Earth-like rotators, the transition from an ice-free dayside to a fully-frozen day-side on synchronous rotators is sharper at higher surface pressures. In general, we find that the increased importance of heat transport on tidally-locked planets results in increasing pN$_2$ causing cooling in more of the parameter space than on fast-rotating Earth-like planets at the same pressures and instellations.

\section{Discussion}
\subsection{Mechanisms of Action on the Climate}\label{sec:mechanisms}

We find that three model components are responsible for the climate's response to different background gas pressures. As was shown and briefly discussed in \autoref{sec:expresults}, the primary contribution to cooling at higher pN$_2$ in our experiments was the reduction in net shortwave heating due to increased reflection by Rayleigh scattering. Our results and those of \citet{Komacek2019} suggest that in some cases, Rayleigh scattering could be sufficient to force planets into a snowball state, similar to the maximum greenhouse phenomenon, where scattering by CO$_2$ overcomes its greenhouse forcing \citep{Kasting1991,Kopparapu2013}. Poleward heat transport does contribute to cooling on planets with Earth-like rotation, but to a lesser extent---its impact on global average temperatures is only apparent when the other major heating and cooling mechanisms are removed. We found that the warming trend in \autoref{fig:pn2grid} could only be reversed by eliminating the water vapor greenhouse effect, either by removing water vapor entirely or by removing pressure broadening. The effect of pressure broadening of H$_2$O is much larger than the effect of broadening CO$_2$ absorption in our models. This is because pressure broadening in PlaSim is an idealized multiplicative factor applied to broadband gas absorptivity, and water is both much more absorptive than CO$_2$ and much more abundant in our models, so the magnitude of amplification for water is much larger than for CO$_2$. A very small amount of cooling from increased pN$_2$ remains when pressure broadening, heat transport, and evaporation have all been removed. When CO$_2$ is also removed, the climate becomes completely insensitive to changes in pN$_2$. 

The amount of warming we find from the inclusion of water vapor is likely to have quantitative differences from what would be found in a more-sophisticated GCM, as the amount and altitude of stratospheric warming by ozone is prescribed in PlaSim. While we checked the effect of removing ozone entirely, we did not explore the impact that choosing different amounts of ozone would have. A more realistic treatment of ozone in atmospheres with varying amounts of background gas might result in different amounts of stratospheric warming, and therefore changes in the efficiency of the tropopause water trap. Similarly, as pressure-broadening in PlaSim is idealized, a more-sophisticated GCM may find different forcings from pressure-broadening of water vapor and CO$_2$.

\begin{figure}
\begin{center}
\includegraphics[width=4in]{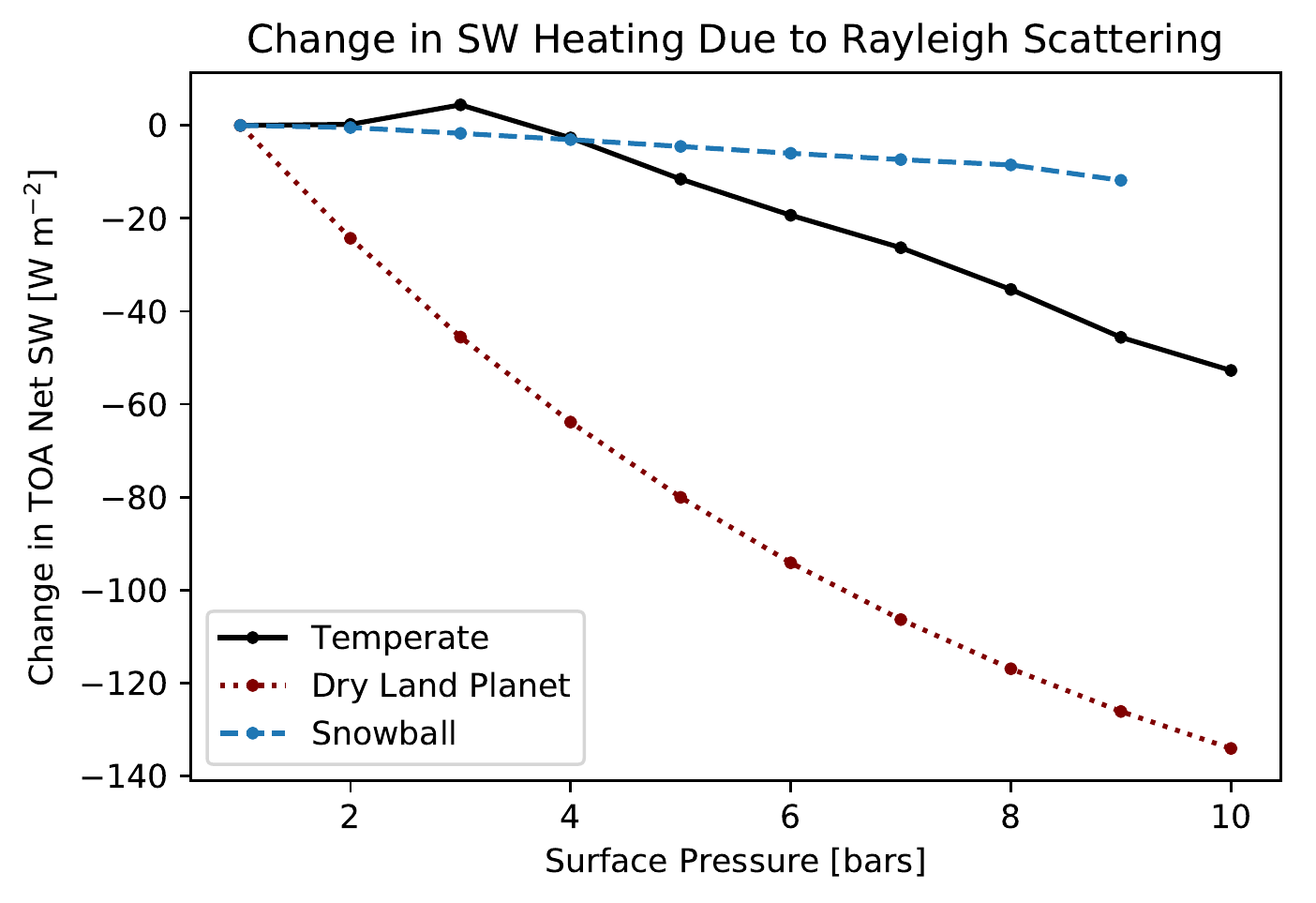}
\end{center}
\caption{Change in shortwave heating due to increased Rayleigh scattering, for temperate and snowball planets at 1300 W m$^{-2}$ and dry land planets at 2000 W m$^{-2}$ with 360 $\mu$bars of CO$_2$. The land planets have uniform gray soil albedos of 0.1. The increase in light reflecting off the atmosphere should be the same for all climates (ignoring cloud effects), but the climate sensitivity of snowball planets to Rayleigh scattering is quite a bit lower than that of either temperate or dry land planets with low soil albedos. This is because the overall top-of-atmosphere albedo of a snowball planet is already high, so an increase in reflection from atmospheric scattering primarily reduces how much light would be reflected off the surface, rather than primarily reducing how much light would be absorbed. This results in a smaller change to the energy budget as measured by net shortwave flux at the top of the atmosphere.}\label{fig:energybudget}
\end{figure}

The relationship between these different mechanisms is very apparent when considering the difference between temperate Earth-like planets and snowball planets.  In contrast to temperate planets, snowball planets demonstrate very low sensitivity to pN$_2$. This low sensitivity is the result of cold temperatures and sea ice imposing limits on evaporation rates, thus limiting the strength of the water vapor positive feedback. The snowball models also appear less sensitive to cooling by Rayleigh scattering because the planet's overall albedo is already very high, due to the highly-reflective surface. An increase in atmospheric albedo can therefore do little to raise the overall albedo, resulting in much smaller changes to the planet's energy budget, as shown in \autoref{fig:energybudget}. In other words, the additional scattered light on a snowball planet is mostly light that would have been reflected anyway, as opposed to a temperate planet or a dry desert planet with a dark surface where that light would have been mostly absorbed. The importance of surface albedo in determining the climate's sensitivity to pN$_2$ suggests that any planet with a dark surface, i.e. planets without extensive ice or snow cover, will undergo strong cooling from increased Rayleigh scattering at high pN$_2$. Land planets which are not completely dry, such as those in \citet{Abe2011}, would exhibit strong cooling from Rayleigh scattering even at snowball-like temperatures, so long as the available water vapor did not produce extensive snow cover.

These results suggest that the effects of changing pN$_2$ can be largely segregated into three different regimes:
\begin{enumerate}
    \item \textbf{High pN$_2$}: Rayleigh scattering dominates, and adding pN$_2$ cools the climate.
    \item \textbf{Wet Surface}: A large surface water reservoir permits strong evaporation at low pN$_2$. The water vapor greenhouse effect dominates due to pressure broadening, and adding pN$_2$ warms the climate.
    \item \textbf{Dry Surface}: Atmospheric water vapor is limited by low evaporation. Rayleigh scattering and heat transport are the dominant effects, and increasing pN$_2$ cools the climate, even at low pN$_2$.
\end{enumerate}
Therefore on planets orbiting Sun-like stars with a limited water vapor greenhouse, either due to evaporation limited by sea ice or an overall lack of surface water, increasing pN$_2$ results in colder climates, while the opposite effect is found on planets with large surface water inventories and enough radiative forcing to drive large amounts of evaporation. At sufficiently high pN$_2$, other details of the climate matter less, as Rayleigh scattering is strong enough to cause cooling regardless of the planet's water inventory. We do not expect a more-realistic layer-by-layer treatment of Rayleigh scattering in PlaSim to change this conclusion, as scattering in a high-pN$_2$ atmosphere would simply occur above where water absorption would occur.

\subsection{Impact of pN$_2$ on Observables}\label{sec:obsv}

Our findings in \autoref{sec:results} suggest that Earth-like climates are highly-sensitive to the amount of background gases present in the atmosphere. However, in the case of triple-bonded diatomic molecules like N$_2$, the absence of prominent absorption lines in the visible or infrared \citep{Lofthus1977} makes it difficult to quantify the abundance of background gases on terrestrial planets with transit spectroscopy \citep{Benneke2012}. Even if the Rayleigh scattering slope as a function of wavelength could be observed in transit spectroscopy, the presence of a cloud deck could limit efforts to constrain the total thickness of an atmosphere \citep{Kaltenegger2007,Benneke2012,Barstow2016}. Observations using reflected light can more-easily probe deeper into the atmosphere \citep[e.g.][]{Snellen2015}, so we investigated the possibility that pN$_2$ might have observable signatures in reflected light.

\begin{figure}
\begin{center}
\includegraphics[width=5in]{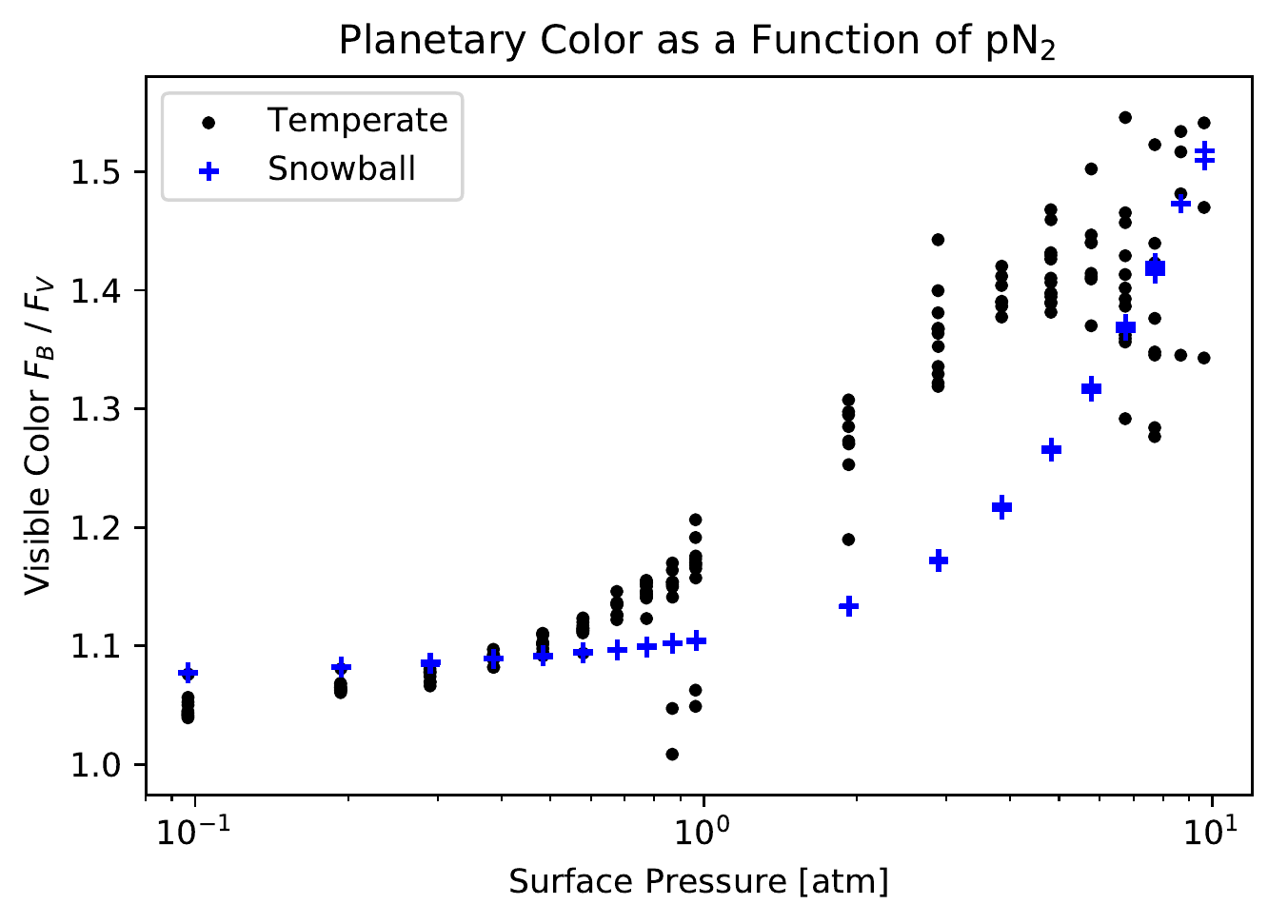}
\end{center}
\caption{B--V photometric colors of Earth-like rotators in reflected light, for each of the models presented in \autoref{fig:pn2grid}. We assume equatorial, zenith viewing geometries consistent with the planet being in secondary eclipse. These fluxes were computed by running SBDART on each column of PlaSim model output.  We find that planets with higher pN$_2$ tend to be bluer due to increased Rayleigh scattering, which may be observable with broadband photometry. However, we caution that we have not investigated confounding factors, and this result therefore cannot be used to infer atmospheric mass and climate from observations. For example, even in the limited observational parameter space represented here, there is significant scatter due to clouds.}\label{fig:colors}
\end{figure}

Our aim here is not to solve the inverse problem in a way that would permit robust and unique retrieval of background gas mass for real exoplanets, but instead to explore how different background gas pressures would affect observables. We therefore limit our analysis to spectra of the models shown in \autoref{fig:pn2grid}, and do not systematically vary cloud fraction, land fraction, photochemistry, or any of the myriad other factors that might confound atmospheric retrievals. The B--V colors of our models are shown in \autoref{fig:colors}. Increasing pN$_2$ results in stronger reflectance at short wavelengths compared to longer wavelengths, resulting in bluer planets. Very strong reflectance by surface ice seems to partially counteract this, potentially because with a reflective surface, red light that is scattered less efficiently than blue light is nonetheless reflected back to space, such that the overall color is relatively unchanged. Conversely, if the surface is dark, as is the case with open ocean, blue light will be scattered back to space while red light will be absorbed at the surface. Because ice is much more reflective than land or ocean, snowball planets therefore appear less blue at moderate background gas partial pressures. Once pN$_2$ is sufficiently high, however, the atmosphere becomes optically-thick to Rayleigh scattering, surface information is obscured, and both temperate and snowball planets appear similarly blue. This applies not only to N$_2$, but to any background gas that causes Rayleigh scattering. For background gases with significant absorption lines such as O$_2$ or CO$_2$, photometric color is probably not the easiest way to constrain atmospheric mass. For gases such as N$_2$ that are otherwise difficult to detect, this finding that thick atmospheres tend to be blue in the absence of color-altering phenomena like photochemical hazes may aid inference efforts.

\subsection{Caveats}

While we believe our results are robust due to qualitative agreement with ExoCAM \citep{Komacek2019} and the 1D models used in \citet{Goldblatt2009} and \citet{Keles2018}, as well as consistency across a diversity of climates, there are nonetheless a number of model caveats that make any quantitative predictions of specific climates suspect. Perhaps most importantly, PlaSim's implementation of Rayleigh scattering assumes that all scattering happens at the atmosphere's bottom layer, which is not physically realistic. The impact on the climate's energy budget for Earth-like surface pressures and lower is negligible, as the atmosphere's exponential density profile means most scattering does indeed happen in the lower layers of the atmosphere, below cloud tops, and absorption by the surface is the primary way visible light enters the climate's energy budget in the regimes we have modeled\citep{pierrehumbertbook}. However, at higher surface pressures, scattering at higher levels of the atmosphere may become important. In models more realistic than PlaSim, atmospheric absorption of visible and near-infrared light by water vapor and oxygen also contributes to the climate's energy budget \citep[e.g.][]{Fujii2017,Kopparapu2017,Way2017}, and scattering in the middle and upper atmosphere may affect that. 

In addition to shortcomings in the scattering parameterization, PlaSim's ozone abundance is prescribed as a normal distribution peaked at 20 km above the surface. This may be appropriate for Earth's atmosphere, but likely incorrect for planets with different N$_2$ and O$_2$ partial pressures. As the height of peak ozone absorption affects stratospheric heating and therefore the height and temperature of the tropopause \citep{Wordsworth2014}, and as this will affect the strength of water-trapping at the tropopause \citep{Zahnle2016}, more work is needed to understand how ozone and photochemistry are affected by different background gas partial pressures, and how their climate feedbacks therefore respond.

While our results show a sharp transition from ice-free states to fully-frozen states at high pressures, we have refrained from examining how the snowball bistability is affected by elevated pN$_2$. This is because the mechanisms by which the climate enters and exits snowball are key to understanding the width of the bistability. An exploration of the changes in the bistability at higher surface pressures should therefore incorporate more-sophisticated GCMs that include key model physics, such as ocean circulation and sea ice drift, and include detailed analysis of how glaciation and deglaciation occur at high pressures. We therefore leave the question of the snowball bistability to future work.

As mentioned in \autoref{sec:obsv}, we have also not considered the role that photochemistry, hazes, aerosols, and differences in cloud formation efficiency might play in thick N$_2$ atmospheres. In addition to affecting observables, these factors would likely also affect the climate's energy budget \citep{Arney2017,Chen2018,Chen2019}. Most of these factors are not feasible to study fully with PlaSim, and so would require further work with a more sophisticated model. 

\begin{figure}
\begin{center}
\includegraphics[width=4in]{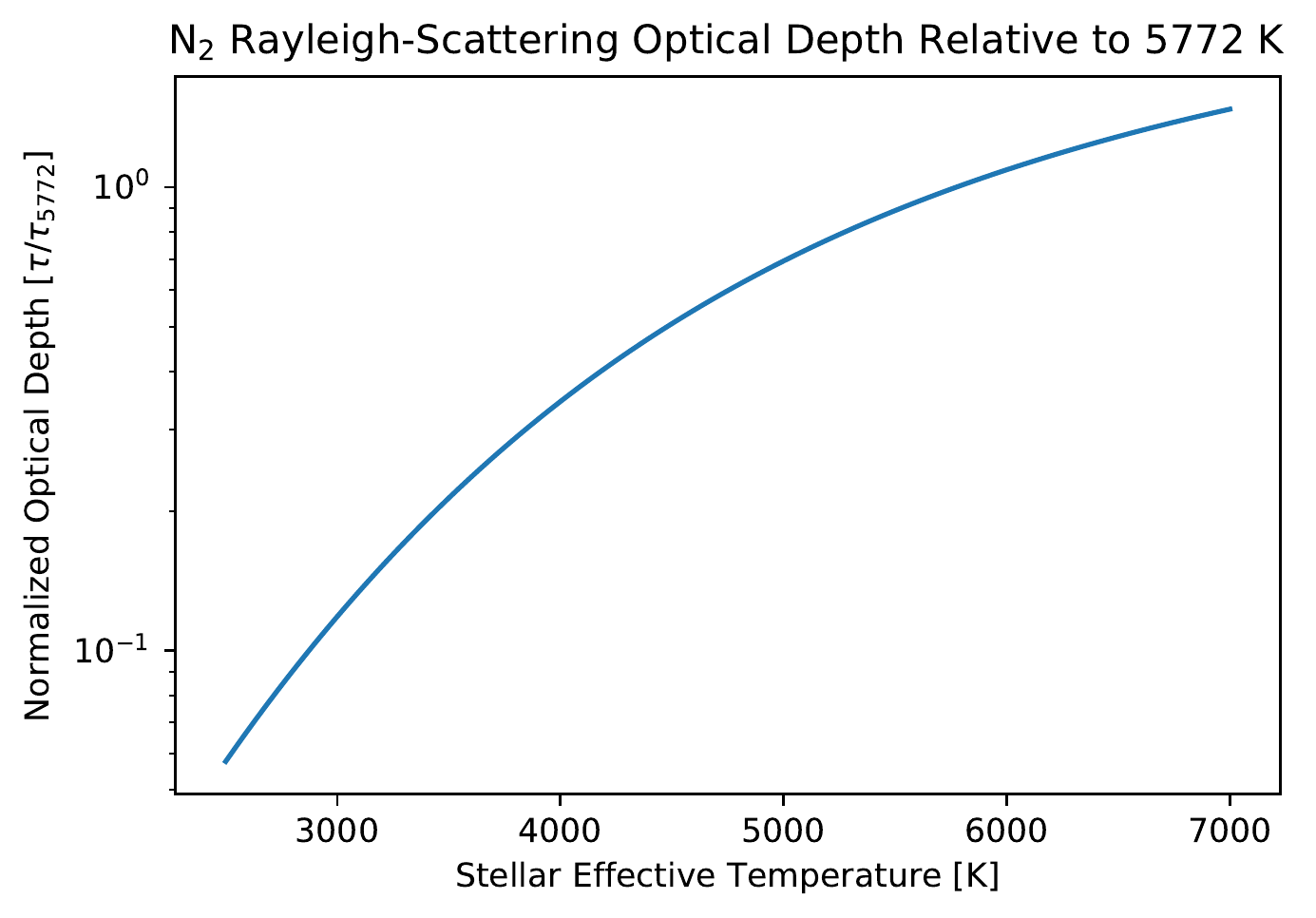}
\end{center}
\caption{Broadband Rayleigh scattering optical depth for a gas like N$_2$ on an Earth-like planet orbiting stars of various effective temperatures, relative to the optical depth at 5772 K (our Sun). This was computed by convolving the scattering cross section with a blackbody spectrum for a given temperature, and integrating over all wavelengths. The Rayleigh scattering cross section is proportional to $\lambda^{-4}$ \citep{Rybicki2004}, so shifting the stellar spectrum's peak to longer wavelengths results in a smaller broadband optical depth. Rayleigh scattering is thus likely to be significantly less important for the climates of planets orbiting cool stars.}\label{fig:rayleighteff}
\end{figure}

We have assumed a solar-like input spectrum in all of our models. One of the major mechanisms by which N$_2$ affects the climate is Rayleigh scattering, which is most efficient at short wavelengths due to a $\lambda^{-4}$ dependence, where $\lambda$ is wavelength\citep{Rybicki2004}. Earth-like planets around lower-mass stars would instead receive most of their light at red and infrared wavelengths, meaning Rayleigh scattering's contribution to the climate's energy budget would be lessened considerably. We have shown this in \autoref{fig:rayleighteff}, in which we calculated how the broadband scattering optical depth for a gas like N$_2$ changes for planets orbiting stars throughout the main sequence. In these climates, we can likely expect cooling through heat transport and warming through pressure broadening to be the dominant ways that N$_2$ affects the climate, as the reduced importance of Rayleigh scattering at high pN$_2$ will reduce the impact of pN$_2$ on shortwave heating. Some work has already been done to explore the impact of changing pN$_2$ on tidally-locked planets orbiting red stars \citep{Turbet2018}, but this has yet to be fully-explored across the habitable zone in a systematic manner, which we leave to future work.

Finally, we did not compute spectra for our tidally-locked models, as almost all real tidally-locked planets will orbit cooler stars with redder spectra \citep{Pierrehumbert2011}. As discussed in the previous paragraph, Rayleigh scattering will be much less important on those planets, so our tidally-locked planets are not representative of the climates likely to be found on real tidally-locked exoplanets. Computing reflected spectra for our tidally-locked models would therefore yield little useful information about the impact of background gas partial pressure on observables. Further work that models tidally-locked planets with appropriately-red input spectra, as in \citet{Turbet2016,Turbet2018,Fauchez2019}, and \citet{Yang2019}, is needed. In particular, future work should aim to produce systematic surveys of the tidally-locked parameter space across the habitable zone.

\section{Conclusion}

We have used a large ensemble of GCMs in which pN$_2$ was varied for a range of instellations and initial conditions, for both Earth-like and tidally-locked rotation, and for both aquaplanet and modern Earth land distributions. We found that the climate's response to pN$_2$ is strong and nonlinear, with increasing pN$_2$ leading to warming in some regimes, and cooling in others. We performed a number of experiments to isolate the mechanisms responsible, and conclude that at low and moderate surface pressures on planets with large surface water reservoirs for evaporation, pressure broadening of the CO$_2$ and H$_2$O absorption lines leads to significant heating. At high surface pressures and in dry climates, Rayleigh scattering dominates over pressure broadening, leading to cooling. On planets with very efficient cooling through horizontal heat transport, such as tidally-locked planets, increased heat transport at high surface pressures further leads to cooling, reducing the potential of pressure broadening to cause net warming. Our results are consistent with those reported in \citet{Komacek2019}, \citet{Goldblatt2009}, and \citet{Keles2018}, and demonstrate the relationship between the various competing mechanisms explored in the literature thus far \citep[e.g.][]{Nakajima1992,Goldblatt2009,Kaspi2015,Zahnle2016}. Finally, we showed that while pN$_2$ may be difficult to quantify with transit spectroscopy, high pN$_2$ will result in planets appearing `bluer' in reflected light, which may eventually be detectable with broadband reflected light photometry through either direct imaging using coronagraphy \citep[e.g.][]{Stark2015}, or through secondary eclipse spectroscopy using differential photometric measurements \citep[e.g.][]{Belu2010}. 

The history of Earth's pN$_2$ is poorly-constrained \citep{Olson2018}, and we have little to no constraints on likely N$_2$ partial pressures on terrestrial exoplanets, particularly those with formation histories and stellar environments different from our own. Even within our own solar system, pN$_2$ varies between terrestrial planets with atmospheres---Venus has over 3 bars of N$_2$ in its atmosphere \citep{Oyama1980}. Our results show however that pN$_2$ is a crucial ingredient in assessing a planet's climate and habitability. We have determined that pN$_2$ will have an impact on observables accessible through direct imaging and secondary eclipse spectroscopy, but more work is needed to understand its role and evolution on planets with both Earth-like and non-Earth-like geophysical and geochemical properties. Furthermore, while increased pN$_2$ has been proposed as a possible resolution to the `faint young Sun paradox' \citep{Johnson2017}, in our sample increased pN$_2$ alone (i.e., without increasing CO$_2$ from its present level) is not able to sufficiently warm an Earth-like planet at Archaean Earth instellation. We have not however fully-explored the range of possible Archaean atmospheres, which likely hosted 10-2500 times the present level of CO$_2$ \citep{Catling2020}. Similarly, studies of the potential habitability and climate of specific exoplanets will need to consider a broad and systematic range of atmospheric compositions and masses, as some recent studies using GCMs \citep[e.g.][]{Turbet2016,DelGenio2018,Turbet2018} have begun to do.

% \begin{outline}
% \1 We robustly show how disagreement on pN$_2$ is reconciled, through competing mechanisms:
%     \2 There is a heating effect from extra pN$_2$ due to pressure-broadening
%     \2 There is a cooling effect due to Rayleigh scattering
%     \2 There is a cooling effect due to heat transport.
% \end{outline}
\section{Acknowledgments}

AP is supported by the Department of Astronomy \& Astrophysics at the University of Toronto, BLF is supported by the Department of Physics at the University of Toronto, KM is supported by the Natural Sciences and Engineering Research Council of Canada, and CL is supported by the Department of Physics at the University of Toronto. Computing resources were provided by the Canadian Institute for Theoretical Astrophysics at the University of Toronto. The authors would like to thank Dorian Abbot for suggestions of existing literature on the role of pN$_2$ in the climate, and the attendees of the Comparative Climatology of Terrestrial Planets III conference in Houston, Texas in August 2018 for an enlightening and thought-provoking discussion about the difficulties in reconstructing the early Earth's pN$_2$ history and evolution. The authors would also like to thank Colin Goldblatt and the anonymous referees for their helpful comments and suggestions, particularly with regards to the model validation in \autoref{fig:comparison}. The modified version of PlaSim used in this study is available on AP's github \citep{gplasim}.

We would furthermore like to acknowledge that our work was performed on land traditionally inhabited by the Wendat, the Anishnaabeg, Haudenosaunee, M\'{e}tis, and the Mississaugas of the Credit First Nation. We also note that this work required supercomputing infrastructure, whose availability was only possible through the mining of precious metals from ecologically-sensitive areas around the world that are traditionally inhabited by Indigenous peoples.

\section{Model Data Availability}\label{sec:data}

The model outputs from this paper are being made available online through \href{https://dataverse.scholarsportal.info/dataset.xhtml?persistentId=doi\%3A10.5683\%2FSP2\%2FLOFVCV}{Dataverse} \citep{pn2data}.

\begin{singlespace}
\bibliographystyle{elsarticle-harv}
% \bibliography{pn2_icarus_rev3_clean_stripped}

\end{singlespace}
\end{document}